%% file: mathieu-arxiv-cleaned.tex
\documentclass[11pt]{article}

\usepackage[dvips,letterpaper]{geometry}

\usepackage{srcltx}

\usepackage{amsmath}
\usepackage{amssymb,amsfonts}

\usepackage{epsfig}
\usepackage[usenames,dvipsnames]{color}
\usepackage[usenames,dvipsnames]{xcolor}
\usepackage{subfigure}

\usepackage{booktabs}

\usepackage{fullpage}
\usepackage{setspace}
\usepackage{flushend}
\usepackage{multicol}

\usepackage{cite}
\usepackage{url}

\usepackage[short,12hr]{datetime}
\usepackage{siunitx}

\usepackage{hyperref}

\usepackage{pst-barcode}

\usepackage{layout}



\newcommand{\ce}{\operatorname{ce}}
\newcommand{\se}{\operatorname{se}}

\title{Elliptic-Cylindrical Wavelets: The Mathieu Wavelets}

\author{%
M.~M.~S.~Lira%
\thanks{M.~M.~S.~Lira is with the Power Systems Digital Laboratory,
Universidade Federal de Pernambuco, Brazil.}
\quad
H.~M.~de~Oliveira%
\thanks{H.~M.~de~Oliveira and R.~J.~Cintra
are 
with the
Signal Processing Group,
Departamento de Estat\'{\i}stica, 
Universidade Federal de Pernambuco,
and were formely with the
Communications Research Group,
Departamento de Eletr\^onica e Sistemas,
Federal University of Pernambuco, Brazil.
E-mail: \{hmo,rjdsc\}@de.ufpe.br
}
\quad
R.~J.~Cintra${}^\dagger$
}

\date{}

\begin{document}

\onehalfspacing

\maketitle

\begin{abstract}
This note introduces a new family of wavelets and a multiresolution analysis, which exploits the relationship between analysing filters and Floquet's solution of Mathieu differential equations. The transfer function of both the detail and the smoothing filter is related to the solution of a Mathieu equation of odd characteristic exponent. The number of notches of these filters can be easily designed. Wavelets derived by this method have potential application in the fields of Optics and Electromagnetism. 
\end{abstract}

\begin{center}
\end{center}

\section{Introduction}

In 1868, the French mathematician \'E. L\'eonard Mathieu introduced a family of differential equations nowadays termed Mathieu equations in his ``\emph{memoir on vibrations of an elliptic membrane}''~\cite{Mat68}.
Mathieu's equation is related to the wave equation for the elliptic cylinder. Mathieu is notably remembered for his discovery of sporadic simple groups~\cite{ConSlo88}. 
This paper is particularly concerned with the canonical form of the Mathieu Equation. 
For $a\in \mathbb{R}$, $q \in \mathbb{C}$, 
the Mathieu Equation is given by
\begin{equation}
\label{eq1}
\frac{\mathrm{d}^2 y}{\mathrm{d}\omega^2}
+
(a - 2 q \cos(2\omega))y
= 0.
\end{equation}
The Mathieu equation is a linear second-order differential equation with periodic coefficients. This equation was shown later to be also related to quantum mechanicals; the parameters $a$ and $q$ denote the energy level and an intensity, respectively. 
For $q=0$ it reduces to the well-known harmonic oscillator, $a$ being the square of the frequency~\cite{McL64}.
The solution of~(\ref{eq1}) is the elliptic-cylindrical harmonic, known as Mathieu functions. In addition to being theoretically fascinating, Mathieu functions are applicable to a wide variety of physical phenomena, e.g., diffraction, amplitude distortion, 
inverted pendulum, 
stability of a floating body, 
radio frequency quadrupole, 
and vibration in a medium with modulated density~\cite{Rub96}. 
They have also long been applied on a broad scope of waveguide problems involving elliptical geometry, including: 
(i) analysis for weak guiding for step index elliptical core optical fibres~\cite{Sha95}, 
(ii) power transport of elliptical waveguides~\cite{HusWur97,Hen98}, 
(iii) evaluating radiated waves of elliptical horn antennas~\cite{Cao97}, 
(iv) elliptical annular microstrip antennas with arbitrary eccentricity~\cite{SunTra93}, 
and (v) scattering by a coated strip~\cite{HolCab92}.

The aim of this paper is to propose a new family of wavelets based on Mathieu differential equations.
Wavelets are a well-known tool for differential equation solving~\cite{Beylkin,Dahmen,Alpert}.
However, in this work, we show another connection between wavelets and differential equations: 
the design of new wavelets from the solution of a differential equation.

\section{Mathieu Equations}

In general, the solutions of~(\ref{eq1}) are not periodic. However, for a given $q$, periodic solutions exist for infinitely many special values (eigenvalues) of $a$. 
For several physically relevant solutions $y$ must be periodic of period $\pi$ or $2\pi$. 
It is also convenient to distinguish even and odd periodic solutions, which are termed Mathieu functions of first kind. One of four simpler types can be considered: Periodic solution ($\pi$ or $2\pi$) symmetry (even or odd). For $q\neq0$, the only periodic solution $y$ corresponding to any characteristic value $a=a_r(q)$ or $a=b_r(q)$ has the following notation:

\noindent
\emph{Even periodic solution}
\begin{equation}
\label{eq2a}
\ce_r(\omega,q)
=
\sum_m
A_{r,m}
\cos m\omega
\qquad
\text{for $a = a_r(q)$},
\tag{2a}
\end{equation}

\noindent
\emph{Odd periodic solution}
\begin{equation}
\label{eq2b}
\se_r(\omega,q)
=
\sum_m
A_{r,m}
\sin m\omega
\qquad
\text{for $a = b_r(q)$},
\tag{2b}
\end{equation}\setcounter{equation}{2}
\!\!%
where the sums are taken over even (respectively odd) values of $m$ if the period of $y$ is $\pi$ (respectively $2\pi$). Given $r$, we denote henceforth $A_{r,m}$ by $A_m$, for short.  
Elliptic cosine and elliptic sine functions are represented by
$\ce$ and $\se$, respectively. 
Interesting relationships are found when $q\to0$, $r\neq0$~\cite{AbraSte68}:
\begin{equation}
\label{eq3}
\lim_{q\to0} \ce_r(\omega,q) = \cos(r\omega), \quad
\lim_{q\to0} \se_r(\omega,q) = \sin(r\omega).
\end{equation}

One of the most powerful results of Mathieu's functions is the Floquet's Theorem~\cite{Flo83}. It states that periodic solutions of~(\ref{eq1}) for any pair $(a, q)$ can be expressed in the form
\begin{equation}
\label{eq5}
\begin{split}
y(\omega)&=
F_\nu(\omega)=
e^{j\nu \omega}P(\omega)
\qquad\text{or}\\
y(\omega)&=
F_\nu(-\omega)=
e^{-j\nu \omega}P(-\omega),
\end{split}
\end{equation}
where $\nu$ is a constant depending on $a$ and $q$ and $P(\cdot)$ is $\pi$-periodic in $\omega$. The constant $\nu$ is called the characteristic exponent. If $\nu$ is an integer, then $F_\nu(\omega)$  and $F_\nu(-\omega)$  are linear dependent solutions. 
Furthermore,  $y(\omega+k\pi) = e^{j\nu k \pi}y(\omega)$ or  
$y(\omega+k\pi) = e^{-j\nu k \pi}y(\omega)$, for the solution  $F_\nu(\omega)$ or $F_\nu(-\omega)$, respectively.  We assume that the pair $(a, q)$ is such that $|\cosh(j\nu\pi)|<1$ so that the solution $y(\omega)$ is bounded on the real axis~\cite{GradRyz65}. 
The general solution of Mathieu's equation ($q\in\mathbb{R}$, $\nu$ non-integer) has the form 
\begin{equation}
\label{eq6}
y(\omega) = c_1 e^{j\nu \omega}P(\omega) + c_2 e^{-j\nu \omega} P(-\omega),
\end{equation}
where $c_1$ and $c_2$ are arbitrary constants.

All bounded solutions ---those of fractional as well as integral order--- are described by an infinite series of harmonic oscillations whose amplitudes decrease with increasing frequency. In the wavelet framework we are basically concerned with even solutions of period $2\pi$. In such cases there exist recurrence relations among the coefficients~\cite{AbraSte68}:
\begin{equation}
\label{eq7}
\begin{split}
(a-1-q)A_1 - qA_3 &= 0, \\
(a-m^2)A_m - q (A_{m-2} + A_{m+2}) &= 0, \quad \text{$m\geq3$, $m$ odd.}
\end{split}
\end{equation}
In the sequel, wavelets are denoted by $\psi(t)$ and scaling functions by $\phi(t)$, with corresponding spectra  $\Psi(\omega)$ and $\Phi(\omega)$, respectively.

\section{Mathieu Wavelets}

Wavelet analysis has matured rapidly over the past years and has been proved to be invaluable for scientists and engineers~\cite{Bul95}. Wavelet transforms have lately gained extensive applications in an amazing number of areas~\cite{Fou95}. 
The equation $\phi(t) = \sqrt2 \sum_{n\in\mathbb{Z}} h_n \phi(2t-n)$, which is known as the \emph{dilation} or \emph{refinement equation}, is the chief relation determining a Multiresolution Analysis (MRA)~\cite{Mall00}. 

\subsection{Two Scale Relation of Scaling Function and Wavelet}

Defining the spectrum of the smoothing filter $\{h_k\}$ by
$
H(\omega) 
\triangleq
\frac{1}{\sqrt2}
\sum_{k\in\mathbb{Z}}
h_k
e^{-j\omega k},
$
the central equations (in the frequency domain) of a Multiresolution analysis are~\cite{Mall89}:
\begin{equation}
\label{eq9}
\Phi(\omega) = H\left(\frac{\omega}{2}\right)\Phi\left(\frac{\omega}{2}\right)
\quad\!\!\text{and}\!\!
\quad 
\Psi(\omega) = G\left(\frac{\omega}{2}\right)\Phi\left(\frac{\omega}{2}\right),
\end{equation}
where 
$G(\omega) \triangleq \frac{1}{\sqrt2} \sum_{k\in\mathbb{Z}} g_k e^{-j\omega k}$, is the transfer function of the detail filter.

The orthogonality condition corresponds to~\cite{Mall89}:
\begin{align}
\label{eq10a}
H(0) = 1\ \text{and}\ H(\pi)=0, \tag{8a}\\
\label{eq10b}
|H(\omega)|^2 + |H(\omega+\pi)|^2 = 1, \tag{8b} \\
\label{eq10c}
H(\omega) = - e^{-j\omega}G^{\ast}(\omega+\pi).\tag{8c}
\end{align}
\setcounter{equation}{8}

\subsection{Filters of a Mathieu MRA}

The subtle liaison between Mathieu's theory and wavelets was found by observing that the classical relationship
\begin{equation}
\Psi(\omega) = 
e^{-j\omega/2}
H^\ast\left( \frac{\omega}{2}-\pi \right) 
\Phi\left( \frac{\omega}{2}\right)
\end{equation}
presents a remarkable similarity to a Floquet's solution of a Mathieu's equation, since $H(\omega)$ is a periodic function.

As a first attempt, the relationship between the wavelet spectrum and the scaling function was put in the form:
\begin{equation}
\label{eq12}
\frac{\Psi(\omega)}{\Phi\left( \frac{\omega}{2}\right)}
=
e^{-j\omega/2}
H^\ast\left( \frac{\omega}{2}-\pi \right).
\end{equation}

Here, on the second member, neither $\nu$ is an integer nor $H(\cdot)$ has a period $\pi$. By an appropriate scaling of this equation, we can rewrite it as
\begin{equation}
\label{eq13}
\frac{\Psi(4\omega)}{\Phi(2\omega)}
=
e^{-j2\omega}
H^\ast(2\omega-\pi).
\end{equation}
Defining a new function   
$Y(\omega) \triangleq {\Psi(4\omega)}/{\Phi(2\omega)}$,
we recognise that it
has a nice interpretation in the wavelet framework. First, we recall that $\Psi(\omega) = G\left(\frac{\omega}{2}\right)\Phi\left(\frac{\omega}{2}\right)$   so that  $\Psi(2\omega) = G(\omega)\Phi(\omega)$. Therefore the function related to Mathieu's equation is exactly $Y(\omega)=G(2\omega)$. Introducing a new variable $z$, which is defined according to  $2z\triangleq 2\omega-\pi$, it follows that  $-Y\left( z + \frac{\pi}{2}\right) = e^{-j2z}H^{\ast}(2z)$.  The characteristic exponent can be adjusted to a particular value $\nu$,
\begin{equation}
\label{eq14}
-e^{-j(\nu-2)z}
Y\left( z + \frac{\pi}{2}\right) 
= 
e^{-j\nu z}H^{\ast}(2z).
\end{equation}
Defining now $P(-z) \triangleq H^\ast(2z)= \sum_{k\in\mathbb{Z}}c_{2k} e^{jz2k}$, where  $c_{2k} \triangleq \frac{1}{\sqrt2}h^\ast_k$, we figure out that the right-side of the above equation represents a Floquet's solution of some differential Mathieu equation. The function $P(\cdot)$ is $\pi$-periodic verifying the initial condition  $P(0) = \frac{1}{\sqrt2}\sum_k h_k = 1$, as expected. The filter coefficients are all assumed to be real. Therefore, there exist a set of parameters $(a_G, q_G)$ such that the auxiliary function 
\begin{equation}
\label{eq17}
y_\nu(z) 
\triangleq
-e^{-j(\nu-2)z} 
Y_\nu\left( z + \frac{\pi}{2} \right)
\end{equation}
is a solution of the following Mathieu equation:
\begin{equation}
\label{eq18}
\frac{\mathrm{d}^2y_\nu}{\mathrm{d}z^2}
+
(a_G - 2 q_G \cos(2z) )
y_\nu = 0,
\end{equation}
subject to $y_\nu(0) = - Y(\pi/2) = -G(\pi) = -1$ and $\cos(\pi \nu) - y_\nu(\pi) = 0$, that is, $y_\nu(\pi) = (-1)^\nu$.

In order to investigate a suitable solution of~(\ref{eq18}), boundary conditions are established for predetermined $a$, $q$. It turns out that when $\nu$ is zero or an integer, $a$ belongs to the set of characteristic values $a_r(q)$. Furthermore, $\nu=r$ is associated with $a_r(q)$. The even ($2\pi$-periodic) solution of such an equation is given by:  
\begin{equation}
\label{eq19}
y_\nu(z) = - \frac{\ce_\nu(z,q)}{\ce_\nu(0,q)}.
\end{equation}
The $Y_\nu(\omega)$ function associated to $y_\nu(z)$ and related to the detail filter of a ``Mathieu MRA'' is thus:
\begin{equation}
Y_\nu(\omega) = G_\nu(2\omega) = 
e^{j(\nu-2)\left( \omega - \frac{\pi}{2}\right)}
\frac{\ce_\nu\left( \omega - \frac{\pi}{2}, q\right)}{\ce_\nu(0,q)}.
\end{equation}
Finally, the transfer function of the detail filter of a Mathieu wavelet is
\begin{equation}
\label{eq21}
G_\nu(\omega) = 
e^{j(\nu-2)\left( \frac{\omega - \pi}{2}\right)}
\frac{\ce_\nu\left( \frac{\omega -\pi}{2}, q\right)}{\ce_\nu(0,q)}.
\end{equation}
The characteristic exponent $\nu$ should be chosen so as to guarantee suitable initial conditions, i.e., $G_\nu(0)=0$ and $G_\nu(\pi)=1$, which are compatible with wavelet filter requirements. Therefore, $\nu$ must be odd. It is interesting to remark that the magnitude of the above transfer function corresponds exactly to the modulus of a elliptic sine~\cite{GradRyz65}:
\begin{equation}
\label{eq22}
|G_\nu(\omega)| = 
\left|
\se_\nu\left( \frac{\omega}{2}, -q \right)/
\ce_\nu(0,q)
\right|.
\end{equation}
The solution for the smoothing filter $H(\cdot)$ can be found out via QMF conditions~\cite{Mall00}, yielding:
\begin{equation}
\label{eq23}
H_\nu(\omega) = 
- e^{-j \nu \frac{\omega}{2}}
\frac{\ce_\nu  \left(  \frac{\omega}{2}, q\right) }{\ce_\nu(0,q)}.
\end{equation}
In this case, we find $H_\nu(\pi)=0$ and 
\begin{equation}
\label{eq24}
|H_\nu(\omega)| = 
\left|
\ce_\nu\left( \frac{\omega}{2}, q \right)/
\ce_\nu(0,q)
\right|.
\end{equation}
Given $q$, the even first-kind Mathieu function with characteristic exponent $\nu$ is given by
$
\ce_\nu(\omega,q) = 
\sum_{l=0}^\infty
A_{2l+1} \cos(2l + 1) \omega
$,
in which $\ce_\nu(0,q) = \sum_{l=0}^\infty A_{2l+1}$.
The $G$ and $H$ filter coefficients of a Mathieu MRA can be expressed in terms of the values $\{ A_{2l+1}\}_{l \in \mathbb{Z}}$  of the Mathieu function as:
\begin{equation}
\label{eq26}
\frac{h_l^\nu}{\sqrt2}
=
-\frac{A_{|2l-\nu|}/2}{\ce_\nu(0,q)}
\quad\text{and}\quad
\frac{g_l^\nu}{\sqrt2}
=
(-1)^l\frac{A_{|2l+\nu-2|}/2}{\ce_\nu(0,q)}.
\end{equation}
It is straightforward to show that 
$h_{-l}^\nu = h_{l+\nu}^\nu$, 
$\forall l>0$. 
The normalising conditions are 
$\frac{1}{\sqrt2}\sum_{k=-\infty}^{\infty}h_k^\nu = -1$ 
and   
$\sum_{k=-\infty}^{\infty}(-1)^k h_k^\nu = 0$.

\begin{figure}
\centering
\subfigure[]{\epsfig{file=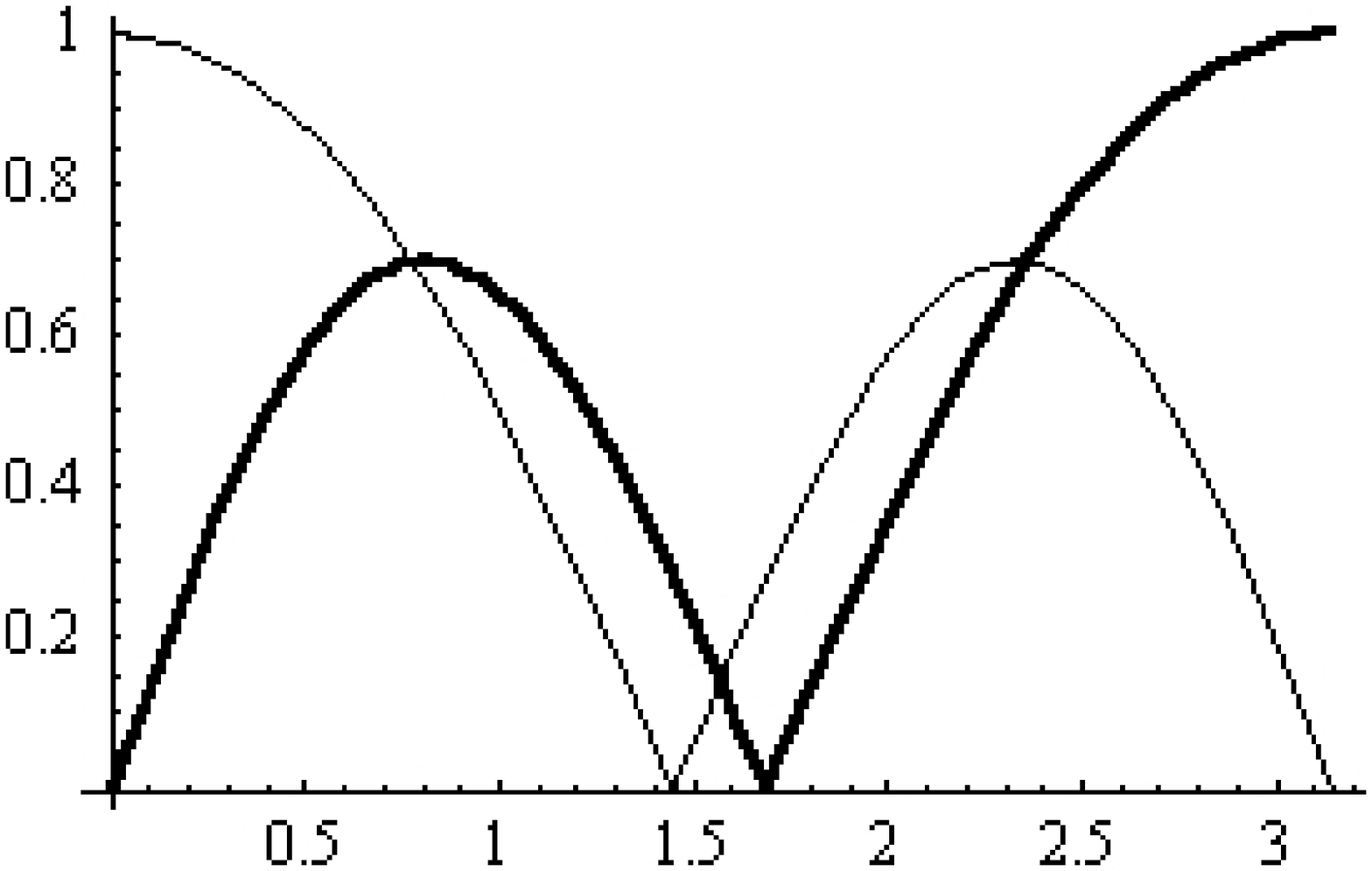,width=.35\linewidth}}
\subfigure[]{\epsfig{file=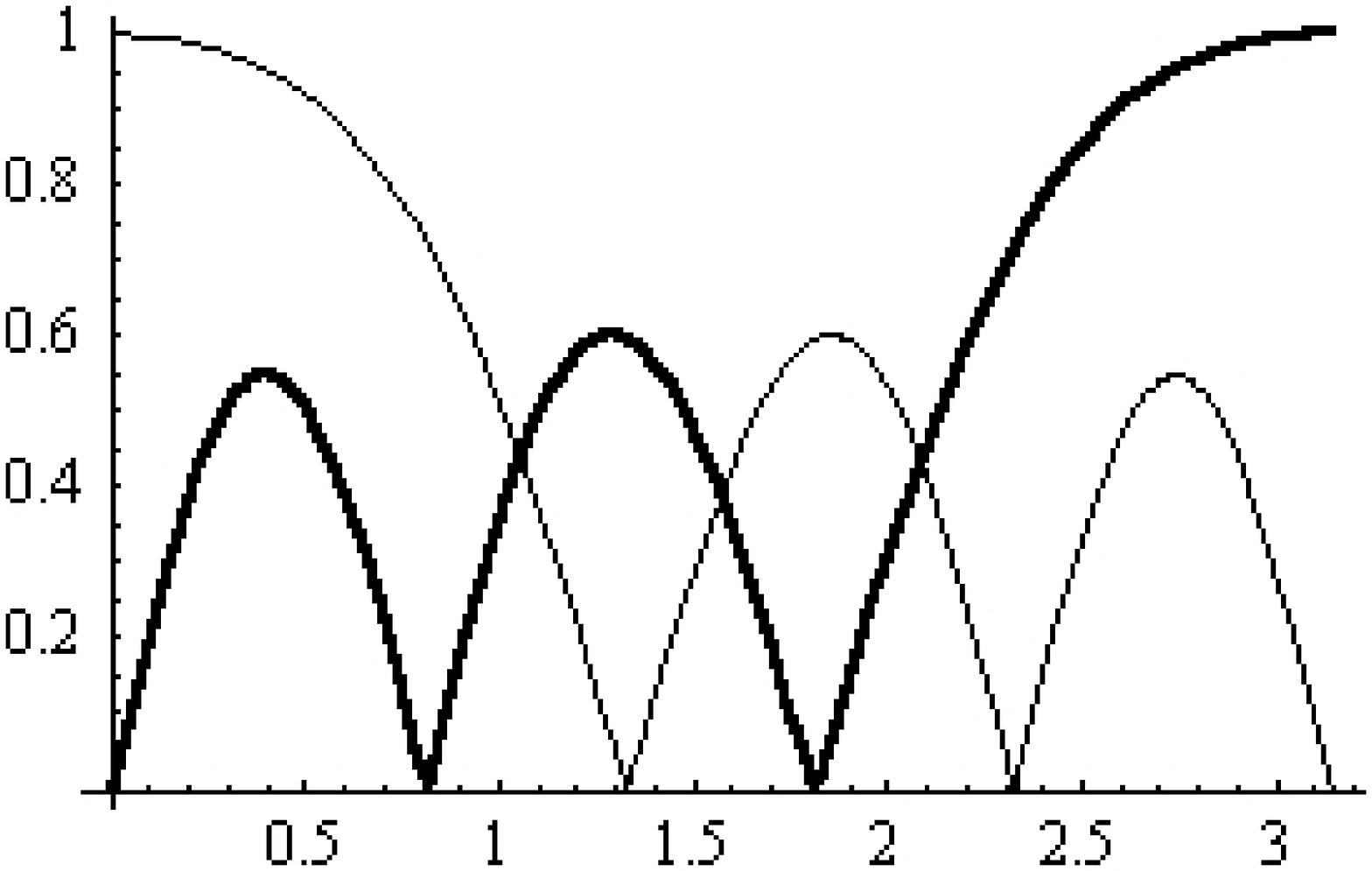,width=.35\linewidth}}
\caption{Magnitude of the transfer function for Mathieu multiresolution analysis filters: smoothing filter $|H_\nu(\omega)|$ (solid line) and detail filter $|G_\nu(\omega)|$ (bold line) for a few Mathieu parameters. 
(a) $\nu=3$, $q=3$, $a = 9.915506290452134$;    
(b) $\nu=5$, $q=15$, $a = 31.957821252172874$.    
}
\label{fig2}
\end{figure}

\begin{figure}
\centering
\subfigure[]{%
\input{fig2a1.latex}
\input{fig2a2.latex}
\input{fig2a3.latex}
}
\subfigure[]{%
\input{fig2b1.latex}
\input{fig2b2.latex}
\input{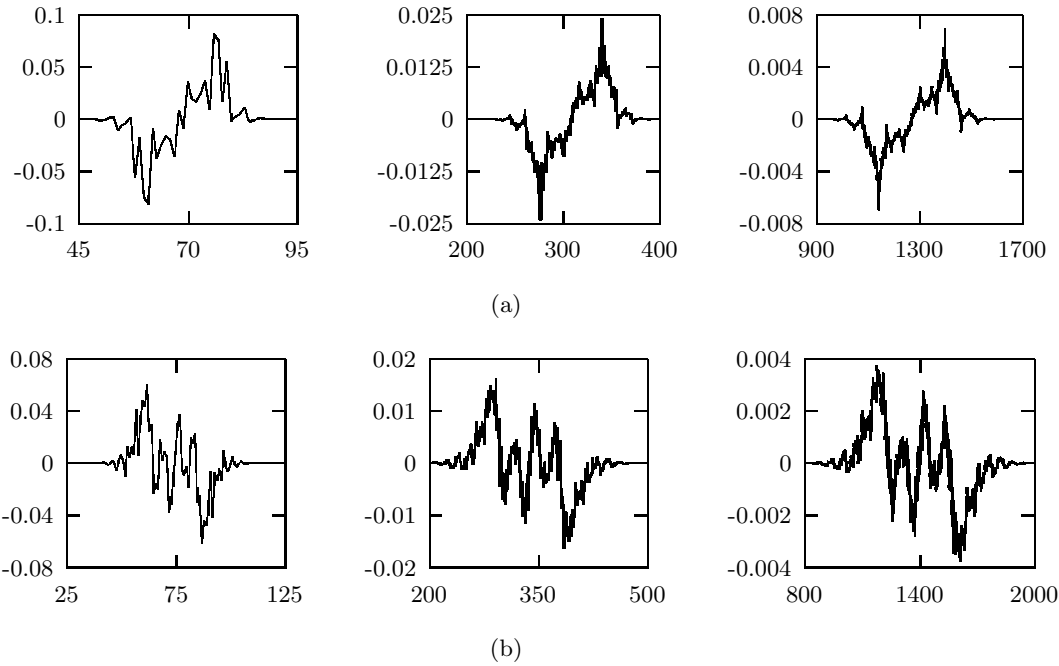}
}

\caption{FIR-Based Approximation of Mathieu Wavelets as the number of iteration increases
(2, 4, and 6 iterations, respectively). 
Filter coefficients holding $|h| <10^{-10}$ were thrown away 
(19 retained coefficients per filter in both cases). 
(a) Mathieu Wavelet with $\nu=3$  and $q=3$ and 
(b) Mathieu Wavelet with $\nu=5$ and $q=15$.}
\label{fig3}
\end{figure}

\section{Examples}

Illustrative examples of filter transfer functions for a Mathieu MRA are shown in Fig.~\ref{fig2}, for $\nu=\text{3}$ and $\text{5}$, and a particular value of $q$ (numerical solution obtained by 5-order Runge-Kutta method). The value of $a$ is  adjusted to an eigenvalue in each case, leading to a periodic solution. Such solutions present a number of $\nu$ zeroes in the interval $|\omega|<\pi$. We observe lowpass behaviour (for the filter $H$) and highpass behaviour (for the filter $G$), as expected. 
Mathieu wavelets can be derived from the lowpass reconstruction filter by the cascade algorithm. %
Infinite Impulse Response filters (IIR) 
should be applied since Mathieu wavelet has no compact support. However a Finite Impulse Response (FIR) approximation can be generated by discarding negligible filter coefficients, say less than $10^{-10}$. 
In Fig.~\ref{fig3}, emerging pattern that progressively looks like the wavelet shape is shown  for some couple of parameter $a$ and~$q$. 
Waveforms were derived using the \textsc{Matlab} wavelet toolbox.
As with many wavelets there is no nice analytical formula for describing Mathieu wavelets. 

\section{Conclusions}

A new and wide family of elliptic-cylindrical wavelets was introduced. It was shown that the transfer functions of the corresponding multiresolution filters are related to Mathieu equation solutions. The magnitude of the detail and smoothing filters corresponds to first-kind Mathieu functions with odd characteristic exponent. The number of zeroes of the highpass $|G(\omega)|$ and lowpass $|H(\omega)|$ filters within the interval $|\omega|<\pi$  can be appropriately designed by choosing the characteristic exponent. This seems to be the first connection found between Mathieu equations and wavelet theory. It opens new perspectives on linking wavelets and solutions of other differential equations (e.g. Associated Legendre functions). 

Although there exist plenty of potential applications for Mathieu Wavelets,
none are presented:
we just disseminate the major ideas,
letting further research to be investigated.
For instance, this new family of wavelets could
be an interesting tool for analysing optical fibres due to 
its ``elliptical'' symmetry. 
They could as well be beneficial when examining molecular dynamics of
charged particles in electromagnetic traps such as Paul trap or the mirror trap for
neutral particles~\cite{NasFoo01,Nieetal01}.

\section*{Acknowledgements}

The authors thank Dr. Harold V. McIntosh from
\emph{Departamento de Aplicaci\'on de Microcomputadoras,
Universidad Aut\'onoma de Puebla}, Mexico for information about Mathieu's functions. They also express their indebtedness to Dr.~R.~M.~Campello de Souza (Federal University of Pernambuco, Brazil) for stimulating comments.

\nocite{*}
\bibliographystyle{ieeetr}%
\bibliography{mathieu-clean}%

\end{document}

%% file: fig2a1.latex
%%%  IMPORTANT REMARK
%%% 
%%%  This figure is in LaTeX format.
%%%  One must \input it in a LaTeX file and compile it.
%
%
%
% GNUPLOT: LaTeX picture
\setlength{\unitlength}{0.240900pt}
\ifx\plotpoint\undefined\newsavebox{\plotpoint}\fi
\begin{picture}(524,450)(0,0)
\font\gnuplot=cmr10 at 10pt
\gnuplot
\footnotesize
\sbox{\plotpoint}{\rule[-0.200pt]{0.400pt}{0.400pt}}%
\put(160.0,82.0){\rule[-0.200pt]{4.818pt}{0.400pt}}
\put(140,82){\makebox(0,0)[r]{-0.1}}
\put(483.0,82.0){\rule[-0.200pt]{4.818pt}{0.400pt}}
\put(160.0,164.0){\rule[-0.200pt]{4.818pt}{0.400pt}}
\put(140,164){\makebox(0,0)[r]{-0.05}}
\put(483.0,164.0){\rule[-0.200pt]{4.818pt}{0.400pt}}
\put(160.0,246.0){\rule[-0.200pt]{4.818pt}{0.400pt}}
\put(140,246){\makebox(0,0)[r]{0}}
\put(483.0,246.0){\rule[-0.200pt]{4.818pt}{0.400pt}}
\put(160.0,328.0){\rule[-0.200pt]{4.818pt}{0.400pt}}
\put(140,328){\makebox(0,0)[r]{0.05}}
\put(483.0,328.0){\rule[-0.200pt]{4.818pt}{0.400pt}}
\put(160.0,410.0){\rule[-0.200pt]{4.818pt}{0.400pt}}
\put(140,410){\makebox(0,0)[r]{0.1}}
\put(483.0,410.0){\rule[-0.200pt]{4.818pt}{0.400pt}}
\put(160.0,82.0){\rule[-0.200pt]{0.400pt}{4.818pt}}
\put(160,41){\makebox(0,0){45}}
\put(160.0,390.0){\rule[-0.200pt]{0.400pt}{4.818pt}}
\put(332.0,82.0){\rule[-0.200pt]{0.400pt}{4.818pt}}
\put(332,41){\makebox(0,0){70}}
\put(332.0,390.0){\rule[-0.200pt]{0.400pt}{4.818pt}}
\put(503.0,82.0){\rule[-0.200pt]{0.400pt}{4.818pt}}
\put(503,41){\makebox(0,0){95}}
\put(503.0,390.0){\rule[-0.200pt]{0.400pt}{4.818pt}}
\put(160.0,82.0){\rule[-0.200pt]{82.629pt}{0.400pt}}
\put(503.0,82.0){\rule[-0.200pt]{0.400pt}{79.015pt}}
\put(160.0,410.0){\rule[-0.200pt]{82.629pt}{0.400pt}}
\put(160.0,82.0){\rule[-0.200pt]{0.400pt}{79.015pt}}
\put(160,246){\usebox{\plotpoint}}
\put(187,244.17){\rule{1.500pt}{0.400pt}}
\multiput(187.00,245.17)(3.887,-2.000){2}{\rule{0.750pt}{0.400pt}}
\put(194,243.67){\rule{1.686pt}{0.400pt}}
\multiput(194.00,243.17)(3.500,1.000){2}{\rule{0.843pt}{0.400pt}}
\multiput(201.00,245.61)(1.355,0.447){3}{\rule{1.033pt}{0.108pt}}
\multiput(201.00,244.17)(4.855,3.000){2}{\rule{0.517pt}{0.400pt}}
\put(208,248.17){\rule{1.500pt}{0.400pt}}
\multiput(208.00,247.17)(3.887,2.000){2}{\rule{0.750pt}{0.400pt}}
\multiput(215.59,244.37)(0.485,-1.637){11}{\rule{0.117pt}{1.357pt}}
\multiput(214.17,247.18)(7.000,-19.183){2}{\rule{0.400pt}{0.679pt}}
\multiput(222.59,228.00)(0.485,0.645){11}{\rule{0.117pt}{0.614pt}}
\multiput(221.17,228.00)(7.000,7.725){2}{\rule{0.400pt}{0.307pt}}
\multiput(229.00,237.60)(0.774,0.468){5}{\rule{0.700pt}{0.113pt}}
\multiput(229.00,236.17)(4.547,4.000){2}{\rule{0.350pt}{0.400pt}}
\multiput(235.59,241.00)(0.485,0.569){11}{\rule{0.117pt}{0.557pt}}
\multiput(234.17,241.00)(7.000,6.844){2}{\rule{0.400pt}{0.279pt}}
\multiput(242.59,226.29)(0.485,-7.129){11}{\rule{0.117pt}{5.471pt}}
\multiput(241.17,237.64)(7.000,-82.644){2}{\rule{0.400pt}{2.736pt}}
\multiput(249.59,155.00)(0.485,4.764){11}{\rule{0.117pt}{3.700pt}}
\multiput(248.17,155.00)(7.000,55.320){2}{\rule{0.400pt}{1.850pt}}
\multiput(256.59,195.05)(0.485,-7.205){11}{\rule{0.117pt}{5.529pt}}
\multiput(255.17,206.53)(7.000,-83.525){2}{\rule{0.400pt}{2.764pt}}
\multiput(263.59,119.98)(0.485,-0.798){11}{\rule{0.117pt}{0.729pt}}
\multiput(262.17,121.49)(7.000,-9.488){2}{\rule{0.400pt}{0.364pt}}
\multiput(270.59,112.00)(0.485,9.036){11}{\rule{0.117pt}{6.900pt}}
\multiput(269.17,112.00)(7.000,104.679){2}{\rule{0.400pt}{3.450pt}}
\multiput(277.59,217.85)(0.482,-4.107){9}{\rule{0.116pt}{3.167pt}}
\multiput(276.17,224.43)(6.000,-39.427){2}{\rule{0.400pt}{1.583pt}}
\multiput(283.59,185.00)(0.485,1.484){11}{\rule{0.117pt}{1.243pt}}
\multiput(282.17,185.00)(7.000,17.420){2}{\rule{0.400pt}{0.621pt}}
\multiput(290.59,205.00)(0.485,1.026){11}{\rule{0.117pt}{0.900pt}}
\multiput(289.17,205.00)(7.000,12.132){2}{\rule{0.400pt}{0.450pt}}
\multiput(297.00,217.93)(0.710,-0.477){7}{\rule{0.660pt}{0.115pt}}
\multiput(297.00,218.17)(5.630,-5.000){2}{\rule{0.330pt}{0.400pt}}
\multiput(304.59,207.18)(0.485,-2.018){11}{\rule{0.117pt}{1.643pt}}
\multiput(303.17,210.59)(7.000,-23.590){2}{\rule{0.400pt}{0.821pt}}
\multiput(311.59,187.00)(0.485,5.527){11}{\rule{0.117pt}{4.271pt}}
\multiput(310.17,187.00)(7.000,64.134){2}{\rule{0.400pt}{2.136pt}}
\multiput(318.59,252.94)(0.485,-2.094){11}{\rule{0.117pt}{1.700pt}}
\multiput(317.17,256.47)(7.000,-24.472){2}{\rule{0.400pt}{0.850pt}}
\multiput(325.59,232.00)(0.485,5.527){11}{\rule{0.117pt}{4.271pt}}
\multiput(324.17,232.00)(7.000,64.134){2}{\rule{0.400pt}{2.136pt}}
\multiput(332.59,297.11)(0.482,-2.389){9}{\rule{0.116pt}{1.900pt}}
\multiput(331.17,301.06)(6.000,-23.056){2}{\rule{0.400pt}{0.950pt}}
\multiput(338.00,276.93)(0.710,-0.477){7}{\rule{0.660pt}{0.115pt}}
\multiput(338.00,277.17)(5.630,-5.000){2}{\rule{0.330pt}{0.400pt}}
\multiput(345.59,273.00)(0.485,1.026){11}{\rule{0.117pt}{0.900pt}}
\multiput(344.17,273.00)(7.000,12.132){2}{\rule{0.400pt}{0.450pt}}
\multiput(352.59,287.00)(0.485,1.484){11}{\rule{0.117pt}{1.243pt}}
\multiput(351.17,287.00)(7.000,17.420){2}{\rule{0.400pt}{0.621pt}}
\multiput(359.59,295.67)(0.485,-3.467){11}{\rule{0.117pt}{2.729pt}}
\multiput(358.17,301.34)(7.000,-40.337){2}{\rule{0.400pt}{1.364pt}}
\multiput(366.59,261.00)(0.485,9.036){11}{\rule{0.117pt}{6.900pt}}
\multiput(365.17,261.00)(7.000,104.679){2}{\rule{0.400pt}{3.450pt}}
\multiput(373.59,376.98)(0.485,-0.798){11}{\rule{0.117pt}{0.729pt}}
\multiput(372.17,378.49)(7.000,-9.488){2}{\rule{0.400pt}{0.364pt}}
\multiput(380.59,342.29)(0.482,-8.537){9}{\rule{0.116pt}{6.433pt}}
\multiput(379.17,355.65)(6.000,-81.647){2}{\rule{0.400pt}{3.217pt}}
\multiput(386.59,274.00)(0.485,4.764){11}{\rule{0.117pt}{3.700pt}}
\multiput(385.17,274.00)(7.000,55.320){2}{\rule{0.400pt}{1.850pt}}
\multiput(393.59,314.29)(0.485,-7.129){11}{\rule{0.117pt}{5.471pt}}
\multiput(392.17,325.64)(7.000,-82.644){2}{\rule{0.400pt}{2.736pt}}
\multiput(400.59,243.00)(0.485,0.569){11}{\rule{0.117pt}{0.557pt}}
\multiput(399.17,243.00)(7.000,6.844){2}{\rule{0.400pt}{0.279pt}}
\multiput(407.00,251.60)(0.920,0.468){5}{\rule{0.800pt}{0.113pt}}
\multiput(407.00,250.17)(5.340,4.000){2}{\rule{0.400pt}{0.400pt}}
\multiput(414.59,255.00)(0.485,0.645){11}{\rule{0.117pt}{0.614pt}}
\multiput(413.17,255.00)(7.000,7.725){2}{\rule{0.400pt}{0.307pt}}
\multiput(421.59,258.37)(0.485,-1.637){11}{\rule{0.117pt}{1.357pt}}
\multiput(420.17,261.18)(7.000,-19.183){2}{\rule{0.400pt}{0.679pt}}
\put(428,242.17){\rule{1.300pt}{0.400pt}}
\multiput(428.00,241.17)(3.302,2.000){2}{\rule{0.650pt}{0.400pt}}
\multiput(434.00,244.61)(1.355,0.447){3}{\rule{1.033pt}{0.108pt}}
\multiput(434.00,243.17)(4.855,3.000){2}{\rule{0.517pt}{0.400pt}}
\put(441,246.67){\rule{1.686pt}{0.400pt}}
\multiput(441.00,246.17)(3.500,1.000){2}{\rule{0.843pt}{0.400pt}}
\put(448,246.17){\rule{1.500pt}{0.400pt}}
\multiput(448.00,247.17)(3.887,-2.000){2}{\rule{0.750pt}{0.400pt}}
\put(160.0,246.0){\rule[-0.200pt]{6.504pt}{0.400pt}}
\put(455.0,246.0){\rule[-0.200pt]{11.563pt}{0.400pt}}
\end{picture}

%% file: fig2a2.latex
%%%  IMPORTANT REMARK
%%% 
%%%  This figure is in LaTeX format.
%%%  One must \input it in a LaTeX file and compile it.
%
%
%
% GNUPLOT: LaTeX picture
\setlength{\unitlength}{0.240900pt}
\ifx\plotpoint\undefined\newsavebox{\plotpoint}\fi
\begin{picture}(524,450)(0,0)
\font\gnuplot=cmr10 at 10pt
\gnuplot
\footnotesize
\sbox{\plotpoint}{\rule[-0.200pt]{0.400pt}{0.400pt}}%
\put(200.0,82.0){\rule[-0.200pt]{4.818pt}{0.400pt}}
\put(180,82){\makebox(0,0)[r]{-0.025}}
\put(483.0,82.0){\rule[-0.200pt]{4.818pt}{0.400pt}}
\put(200.0,164.0){\rule[-0.200pt]{4.818pt}{0.400pt}}
\put(180,164){\makebox(0,0)[r]{-0.0125}}
\put(483.0,164.0){\rule[-0.200pt]{4.818pt}{0.400pt}}
\put(200.0,246.0){\rule[-0.200pt]{4.818pt}{0.400pt}}
\put(180,246){\makebox(0,0)[r]{0}}
\put(483.0,246.0){\rule[-0.200pt]{4.818pt}{0.400pt}}
\put(200.0,328.0){\rule[-0.200pt]{4.818pt}{0.400pt}}
\put(180,328){\makebox(0,0)[r]{0.0125}}
\put(483.0,328.0){\rule[-0.200pt]{4.818pt}{0.400pt}}
\put(200.0,410.0){\rule[-0.200pt]{4.818pt}{0.400pt}}
\put(180,410){\makebox(0,0)[r]{0.025}}
\put(483.0,410.0){\rule[-0.200pt]{4.818pt}{0.400pt}}
\put(200.0,82.0){\rule[-0.200pt]{0.400pt}{4.818pt}}
\put(200,41){\makebox(0,0){200}}
\put(200.0,390.0){\rule[-0.200pt]{0.400pt}{4.818pt}}
\put(351.0,82.0){\rule[-0.200pt]{0.400pt}{4.818pt}}
\put(351,41){\makebox(0,0){300}}
\put(351.0,390.0){\rule[-0.200pt]{0.400pt}{4.818pt}}
\put(503.0,82.0){\rule[-0.200pt]{0.400pt}{4.818pt}}
\put(503,41){\makebox(0,0){400}}
\put(503.0,390.0){\rule[-0.200pt]{0.400pt}{4.818pt}}
\put(200.0,82.0){\rule[-0.200pt]{72.993pt}{0.400pt}}
\put(503.0,82.0){\rule[-0.200pt]{0.400pt}{79.015pt}}
\put(200.0,410.0){\rule[-0.200pt]{72.993pt}{0.400pt}}
\put(200.0,82.0){\rule[-0.200pt]{0.400pt}{79.015pt}}
\put(200,246){\usebox{\plotpoint}}
\put(241,245.67){\rule{0.241pt}{0.400pt}}
\multiput(241.00,245.17)(0.500,1.000){2}{\rule{0.120pt}{0.400pt}}
\put(200.0,246.0){\rule[-0.200pt]{9.877pt}{0.400pt}}
\put(243.67,245){\rule{0.400pt}{0.482pt}}
\multiput(243.17,246.00)(1.000,-1.000){2}{\rule{0.400pt}{0.241pt}}
\put(245,244.67){\rule{0.482pt}{0.400pt}}
\multiput(245.00,244.17)(1.000,1.000){2}{\rule{0.241pt}{0.400pt}}
\put(247,244.67){\rule{0.241pt}{0.400pt}}
\multiput(247.00,245.17)(0.500,-1.000){2}{\rule{0.120pt}{0.400pt}}
\put(242.0,247.0){\rule[-0.200pt]{0.482pt}{0.400pt}}
\put(252,244.67){\rule{0.241pt}{0.400pt}}
\multiput(252.00,244.17)(0.500,1.000){2}{\rule{0.120pt}{0.400pt}}
\put(253,244.17){\rule{0.482pt}{0.400pt}}
\multiput(253.00,245.17)(1.000,-2.000){2}{\rule{0.241pt}{0.400pt}}
\put(248.0,245.0){\rule[-0.200pt]{0.964pt}{0.400pt}}
\put(256,244.17){\rule{0.482pt}{0.400pt}}
\multiput(256.00,243.17)(1.000,2.000){2}{\rule{0.241pt}{0.400pt}}
\put(258,244.67){\rule{0.241pt}{0.400pt}}
\multiput(258.00,245.17)(0.500,-1.000){2}{\rule{0.120pt}{0.400pt}}
\put(259,244.67){\rule{0.482pt}{0.400pt}}
\multiput(259.00,244.17)(1.000,1.000){2}{\rule{0.241pt}{0.400pt}}
\put(255.0,244.0){\usebox{\plotpoint}}
\put(262,246.17){\rule{0.482pt}{0.400pt}}
\multiput(262.00,245.17)(1.000,2.000){2}{\rule{0.241pt}{0.400pt}}
\put(263.67,246){\rule{0.400pt}{0.482pt}}
\multiput(263.17,247.00)(1.000,-1.000){2}{\rule{0.400pt}{0.241pt}}
\put(265.17,246){\rule{0.400pt}{1.300pt}}
\multiput(264.17,246.00)(2.000,3.302){2}{\rule{0.400pt}{0.650pt}}
\put(267,251.67){\rule{0.241pt}{0.400pt}}
\multiput(267.00,251.17)(0.500,1.000){2}{\rule{0.120pt}{0.400pt}}
\put(268.17,238){\rule{0.400pt}{3.100pt}}
\multiput(267.17,246.57)(2.000,-8.566){2}{\rule{0.400pt}{1.550pt}}
\put(269.67,238){\rule{0.400pt}{1.927pt}}
\multiput(269.17,238.00)(1.000,4.000){2}{\rule{0.400pt}{0.964pt}}
\put(271.17,238){\rule{0.400pt}{1.700pt}}
\multiput(270.17,242.47)(2.000,-4.472){2}{\rule{0.400pt}{0.850pt}}
\put(261.0,246.0){\usebox{\plotpoint}}
\put(274,236.17){\rule{0.482pt}{0.400pt}}
\multiput(274.00,237.17)(1.000,-2.000){2}{\rule{0.241pt}{0.400pt}}
\put(275.67,236){\rule{0.400pt}{1.204pt}}
\multiput(275.17,236.00)(1.000,2.500){2}{\rule{0.400pt}{0.602pt}}
\put(277.17,229){\rule{0.400pt}{2.500pt}}
\multiput(276.17,235.81)(2.000,-6.811){2}{\rule{0.400pt}{1.250pt}}
\put(279,228.67){\rule{0.241pt}{0.400pt}}
\multiput(279.00,228.17)(0.500,1.000){2}{\rule{0.120pt}{0.400pt}}
\put(280.17,230){\rule{0.400pt}{1.900pt}}
\multiput(279.17,230.00)(2.000,5.056){2}{\rule{0.400pt}{0.950pt}}
\put(281.67,233){\rule{0.400pt}{1.445pt}}
\multiput(281.17,236.00)(1.000,-3.000){2}{\rule{0.400pt}{0.723pt}}
\put(283.17,233){\rule{0.400pt}{1.500pt}}
\multiput(282.17,233.00)(2.000,3.887){2}{\rule{0.400pt}{0.750pt}}
\put(285,238.67){\rule{0.241pt}{0.400pt}}
\multiput(285.00,239.17)(0.500,-1.000){2}{\rule{0.120pt}{0.400pt}}
\put(286.17,239){\rule{0.400pt}{1.100pt}}
\multiput(285.17,239.00)(2.000,2.717){2}{\rule{0.400pt}{0.550pt}}
\put(287.67,238){\rule{0.400pt}{1.445pt}}
\multiput(287.17,241.00)(1.000,-3.000){2}{\rule{0.400pt}{0.723pt}}
\put(289.17,238){\rule{0.400pt}{3.700pt}}
\multiput(288.17,238.00)(2.000,10.320){2}{\rule{0.400pt}{1.850pt}}
\put(290.67,256){\rule{0.400pt}{1.204pt}}
\multiput(290.17,256.00)(1.000,2.500){2}{\rule{0.400pt}{0.602pt}}
\put(292.17,199){\rule{0.400pt}{12.500pt}}
\multiput(291.17,235.06)(2.000,-36.056){2}{\rule{0.400pt}{6.250pt}}
\put(293.67,199){\rule{0.400pt}{8.191pt}}
\multiput(293.17,199.00)(1.000,17.000){2}{\rule{0.400pt}{4.095pt}}
\put(295.17,197){\rule{0.400pt}{7.300pt}}
\multiput(294.17,217.85)(2.000,-20.848){2}{\rule{0.400pt}{3.650pt}}
\put(296.67,194){\rule{0.400pt}{0.723pt}}
\multiput(296.17,195.50)(1.000,-1.500){2}{\rule{0.400pt}{0.361pt}}
\put(298.17,194){\rule{0.400pt}{1.500pt}}
\multiput(297.17,194.00)(2.000,3.887){2}{\rule{0.400pt}{0.750pt}}
\put(300.17,201){\rule{0.400pt}{2.100pt}}
\multiput(299.17,201.00)(2.000,5.641){2}{\rule{0.400pt}{1.050pt}}
\put(301.67,181){\rule{0.400pt}{7.227pt}}
\multiput(301.17,196.00)(1.000,-15.000){2}{\rule{0.400pt}{3.613pt}}
\put(303.17,181){\rule{0.400pt}{2.500pt}}
\multiput(302.17,181.00)(2.000,6.811){2}{\rule{0.400pt}{1.250pt}}
\put(304.67,162){\rule{0.400pt}{7.468pt}}
\multiput(304.17,177.50)(1.000,-15.500){2}{\rule{0.400pt}{3.734pt}}
\put(306.17,162){\rule{0.400pt}{2.100pt}}
\multiput(305.17,162.00)(2.000,5.641){2}{\rule{0.400pt}{1.050pt}}
\put(307.67,172){\rule{0.400pt}{2.168pt}}
\multiput(307.17,172.00)(1.000,4.500){2}{\rule{0.400pt}{1.084pt}}
\put(309,179.17){\rule{0.482pt}{0.400pt}}
\multiput(309.00,180.17)(1.000,-2.000){2}{\rule{0.241pt}{0.400pt}}
\put(310.67,130){\rule{0.400pt}{11.804pt}}
\multiput(310.17,154.50)(1.000,-24.500){2}{\rule{0.400pt}{5.902pt}}
\put(312.17,130){\rule{0.400pt}{8.900pt}}
\multiput(311.17,130.00)(2.000,25.528){2}{\rule{0.400pt}{4.450pt}}
\put(313.67,89){\rule{0.400pt}{20.476pt}}
\multiput(313.17,131.50)(1.000,-42.500){2}{\rule{0.400pt}{10.238pt}}
\put(315,87.67){\rule{0.482pt}{0.400pt}}
\multiput(315.00,88.17)(1.000,-1.000){2}{\rule{0.241pt}{0.400pt}}
\put(316.67,88){\rule{0.400pt}{21.681pt}}
\multiput(316.17,88.00)(1.000,45.000){2}{\rule{0.400pt}{10.840pt}}
\put(318.17,134){\rule{0.400pt}{8.900pt}}
\multiput(317.17,159.53)(2.000,-25.528){2}{\rule{0.400pt}{4.450pt}}
\put(319.67,134){\rule{0.400pt}{9.877pt}}
\multiput(319.17,134.00)(1.000,20.500){2}{\rule{0.400pt}{4.938pt}}
\put(321,175.17){\rule{0.482pt}{0.400pt}}
\multiput(321.00,174.17)(1.000,2.000){2}{\rule{0.241pt}{0.400pt}}
\put(322.67,177){\rule{0.400pt}{3.132pt}}
\multiput(322.17,177.00)(1.000,6.500){2}{\rule{0.400pt}{1.566pt}}
\put(324.17,162){\rule{0.400pt}{5.700pt}}
\multiput(323.17,178.17)(2.000,-16.169){2}{\rule{0.400pt}{2.850pt}}
\put(325.67,162){\rule{0.400pt}{15.418pt}}
\multiput(325.17,162.00)(1.000,32.000){2}{\rule{0.400pt}{7.709pt}}
\put(327.17,217){\rule{0.400pt}{1.900pt}}
\multiput(326.17,222.06)(2.000,-5.056){2}{\rule{0.400pt}{0.950pt}}
\put(328.67,199){\rule{0.400pt}{4.336pt}}
\multiput(328.17,208.00)(1.000,-9.000){2}{\rule{0.400pt}{2.168pt}}
\put(330.17,199){\rule{0.400pt}{2.700pt}}
\multiput(329.17,199.00)(2.000,7.396){2}{\rule{0.400pt}{1.350pt}}
\put(331.67,186){\rule{0.400pt}{6.263pt}}
\multiput(331.17,199.00)(1.000,-13.000){2}{\rule{0.400pt}{3.132pt}}
\put(333,186.17){\rule{0.482pt}{0.400pt}}
\multiput(333.00,185.17)(1.000,2.000){2}{\rule{0.241pt}{0.400pt}}
\put(334.67,188){\rule{0.400pt}{5.541pt}}
\multiput(334.17,188.00)(1.000,11.500){2}{\rule{0.400pt}{2.770pt}}
\put(336.17,197){\rule{0.400pt}{2.900pt}}
\multiput(335.17,204.98)(2.000,-7.981){2}{\rule{0.400pt}{1.450pt}}
\put(337.67,197){\rule{0.400pt}{4.577pt}}
\multiput(337.17,197.00)(1.000,9.500){2}{\rule{0.400pt}{2.289pt}}
\put(339,214.67){\rule{0.482pt}{0.400pt}}
\multiput(339.00,215.17)(1.000,-1.000){2}{\rule{0.241pt}{0.400pt}}
\put(340.67,213){\rule{0.400pt}{0.482pt}}
\multiput(340.17,214.00)(1.000,-1.000){2}{\rule{0.400pt}{0.241pt}}
\put(342.17,210){\rule{0.400pt}{0.700pt}}
\multiput(341.17,211.55)(2.000,-1.547){2}{\rule{0.400pt}{0.350pt}}
\put(343.67,210){\rule{0.400pt}{2.891pt}}
\multiput(343.17,210.00)(1.000,6.000){2}{\rule{0.400pt}{1.445pt}}
\put(273.0,238.0){\usebox{\plotpoint}}
\put(346.67,200){\rule{0.400pt}{5.300pt}}
\multiput(346.17,211.00)(1.000,-11.000){2}{\rule{0.400pt}{2.650pt}}
\put(348.17,200){\rule{0.400pt}{3.100pt}}
\multiput(347.17,200.00)(2.000,8.566){2}{\rule{0.400pt}{1.550pt}}
\put(349.67,191){\rule{0.400pt}{5.782pt}}
\multiput(349.17,203.00)(1.000,-12.000){2}{\rule{0.400pt}{2.891pt}}
\put(351.17,187){\rule{0.400pt}{0.900pt}}
\multiput(350.17,189.13)(2.000,-2.132){2}{\rule{0.400pt}{0.450pt}}
\put(353.17,187){\rule{0.400pt}{8.500pt}}
\multiput(352.17,187.00)(2.000,24.358){2}{\rule{0.400pt}{4.250pt}}
\put(354.67,211){\rule{0.400pt}{4.336pt}}
\multiput(354.17,220.00)(1.000,-9.000){2}{\rule{0.400pt}{2.168pt}}
\put(356.17,211){\rule{0.400pt}{2.300pt}}
\multiput(355.17,211.00)(2.000,6.226){2}{\rule{0.400pt}{1.150pt}}
\put(357.67,222){\rule{0.400pt}{0.482pt}}
\multiput(357.17,222.00)(1.000,1.000){2}{\rule{0.400pt}{0.241pt}}
\put(359.17,224){\rule{0.400pt}{2.500pt}}
\multiput(358.17,224.00)(2.000,6.811){2}{\rule{0.400pt}{1.250pt}}
\put(360.67,220){\rule{0.400pt}{3.854pt}}
\multiput(360.17,228.00)(1.000,-8.000){2}{\rule{0.400pt}{1.927pt}}
\put(362.17,220){\rule{0.400pt}{6.300pt}}
\multiput(361.17,220.00)(2.000,17.924){2}{\rule{0.400pt}{3.150pt}}
\put(363.67,241){\rule{0.400pt}{2.409pt}}
\multiput(363.17,246.00)(1.000,-5.000){2}{\rule{0.400pt}{1.204pt}}
\put(365.17,241){\rule{0.400pt}{6.300pt}}
\multiput(364.17,241.00)(2.000,17.924){2}{\rule{0.400pt}{3.150pt}}
\put(366.67,256){\rule{0.400pt}{3.854pt}}
\multiput(366.17,264.00)(1.000,-8.000){2}{\rule{0.400pt}{1.927pt}}
\put(368.17,256){\rule{0.400pt}{2.500pt}}
\multiput(367.17,256.00)(2.000,6.811){2}{\rule{0.400pt}{1.250pt}}
\put(369.67,268){\rule{0.400pt}{0.482pt}}
\multiput(369.17,268.00)(1.000,1.000){2}{\rule{0.400pt}{0.241pt}}
\put(371.17,270){\rule{0.400pt}{2.300pt}}
\multiput(370.17,270.00)(2.000,6.226){2}{\rule{0.400pt}{1.150pt}}
\put(372.67,263){\rule{0.400pt}{4.336pt}}
\multiput(372.17,272.00)(1.000,-9.000){2}{\rule{0.400pt}{2.168pt}}
\put(374.17,263){\rule{0.400pt}{8.500pt}}
\multiput(373.17,263.00)(2.000,24.358){2}{\rule{0.400pt}{4.250pt}}
\put(375.67,301){\rule{0.400pt}{0.964pt}}
\multiput(375.17,303.00)(1.000,-2.000){2}{\rule{0.400pt}{0.482pt}}
\put(377.17,277){\rule{0.400pt}{4.900pt}}
\multiput(376.17,290.83)(2.000,-13.830){2}{\rule{0.400pt}{2.450pt}}
\put(378.67,277){\rule{0.400pt}{3.614pt}}
\multiput(378.17,277.00)(1.000,7.500){2}{\rule{0.400pt}{1.807pt}}
\put(380.17,270){\rule{0.400pt}{4.500pt}}
\multiput(379.17,282.66)(2.000,-12.660){2}{\rule{0.400pt}{2.250pt}}
\put(345.0,222.0){\rule[-0.200pt]{0.482pt}{0.400pt}}
\put(383.17,270){\rule{0.400pt}{2.500pt}}
\multiput(382.17,270.00)(2.000,6.811){2}{\rule{0.400pt}{1.250pt}}
\put(384.67,279){\rule{0.400pt}{0.723pt}}
\multiput(384.17,280.50)(1.000,-1.500){2}{\rule{0.400pt}{0.361pt}}
\put(386,277.17){\rule{0.482pt}{0.400pt}}
\multiput(386.00,278.17)(1.000,-2.000){2}{\rule{0.241pt}{0.400pt}}
\put(388,275.67){\rule{0.241pt}{0.400pt}}
\multiput(388.00,276.17)(0.500,-1.000){2}{\rule{0.120pt}{0.400pt}}
\put(389.17,276){\rule{0.400pt}{3.900pt}}
\multiput(388.17,276.00)(2.000,10.905){2}{\rule{0.400pt}{1.950pt}}
\put(390.67,281){\rule{0.400pt}{3.373pt}}
\multiput(390.17,288.00)(1.000,-7.000){2}{\rule{0.400pt}{1.686pt}}
\put(392.17,281){\rule{0.400pt}{4.700pt}}
\multiput(391.17,281.00)(2.000,13.245){2}{\rule{0.400pt}{2.350pt}}
\put(393.67,304){\rule{0.400pt}{0.482pt}}
\multiput(393.17,304.00)(1.000,1.000){2}{\rule{0.400pt}{0.241pt}}
\put(395.17,280){\rule{0.400pt}{5.300pt}}
\multiput(394.17,295.00)(2.000,-15.000){2}{\rule{0.400pt}{2.650pt}}
\put(396.67,280){\rule{0.400pt}{3.132pt}}
\multiput(396.17,280.00)(1.000,6.500){2}{\rule{0.400pt}{1.566pt}}
\put(398.17,275){\rule{0.400pt}{3.700pt}}
\multiput(397.17,285.32)(2.000,-10.320){2}{\rule{0.400pt}{1.850pt}}
\put(399.67,266){\rule{0.400pt}{2.168pt}}
\multiput(399.17,270.50)(1.000,-4.500){2}{\rule{0.400pt}{1.084pt}}
\put(401.17,266){\rule{0.400pt}{12.900pt}}
\multiput(400.17,266.00)(2.000,37.225){2}{\rule{0.400pt}{6.450pt}}
\put(403.17,302){\rule{0.400pt}{5.700pt}}
\multiput(402.17,318.17)(2.000,-16.169){2}{\rule{0.400pt}{2.850pt}}
\put(404.67,302){\rule{0.400pt}{3.132pt}}
\multiput(404.17,302.00)(1.000,6.500){2}{\rule{0.400pt}{1.566pt}}
\put(406,315.17){\rule{0.482pt}{0.400pt}}
\multiput(406.00,314.17)(1.000,2.000){2}{\rule{0.241pt}{0.400pt}}
\put(407.67,317){\rule{0.400pt}{9.877pt}}
\multiput(407.17,317.00)(1.000,20.500){2}{\rule{0.400pt}{4.938pt}}
\put(409.17,314){\rule{0.400pt}{8.900pt}}
\multiput(408.17,339.53)(2.000,-25.528){2}{\rule{0.400pt}{4.450pt}}
\put(410.67,314){\rule{0.400pt}{21.681pt}}
\multiput(410.17,314.00)(1.000,45.000){2}{\rule{0.400pt}{10.840pt}}
\put(412,402.67){\rule{0.482pt}{0.400pt}}
\multiput(412.00,403.17)(1.000,-1.000){2}{\rule{0.241pt}{0.400pt}}
\put(413.67,318){\rule{0.400pt}{20.476pt}}
\multiput(413.17,360.50)(1.000,-42.500){2}{\rule{0.400pt}{10.238pt}}
\put(415.17,318){\rule{0.400pt}{8.900pt}}
\multiput(414.17,318.00)(2.000,25.528){2}{\rule{0.400pt}{4.450pt}}
\put(416.67,313){\rule{0.400pt}{11.804pt}}
\multiput(416.17,337.50)(1.000,-24.500){2}{\rule{0.400pt}{5.902pt}}
\put(418,311.17){\rule{0.482pt}{0.400pt}}
\multiput(418.00,312.17)(1.000,-2.000){2}{\rule{0.241pt}{0.400pt}}
\put(419.67,311){\rule{0.400pt}{2.168pt}}
\multiput(419.17,311.00)(1.000,4.500){2}{\rule{0.400pt}{1.084pt}}
\put(421.17,320){\rule{0.400pt}{2.100pt}}
\multiput(420.17,320.00)(2.000,5.641){2}{\rule{0.400pt}{1.050pt}}
\put(422.67,299){\rule{0.400pt}{7.468pt}}
\multiput(422.17,314.50)(1.000,-15.500){2}{\rule{0.400pt}{3.734pt}}
\put(424.17,299){\rule{0.400pt}{2.500pt}}
\multiput(423.17,299.00)(2.000,6.811){2}{\rule{0.400pt}{1.250pt}}
\put(425.67,281){\rule{0.400pt}{7.227pt}}
\multiput(425.17,296.00)(1.000,-15.000){2}{\rule{0.400pt}{3.613pt}}
\put(427.17,281){\rule{0.400pt}{2.100pt}}
\multiput(426.17,281.00)(2.000,5.641){2}{\rule{0.400pt}{1.050pt}}
\put(428.67,291){\rule{0.400pt}{1.686pt}}
\multiput(428.17,291.00)(1.000,3.500){2}{\rule{0.400pt}{0.843pt}}
\put(430.17,295){\rule{0.400pt}{0.700pt}}
\multiput(429.17,296.55)(2.000,-1.547){2}{\rule{0.400pt}{0.350pt}}
\put(431.67,259){\rule{0.400pt}{8.672pt}}
\multiput(431.17,277.00)(1.000,-18.000){2}{\rule{0.400pt}{4.336pt}}
\put(433.17,259){\rule{0.400pt}{6.900pt}}
\multiput(432.17,259.00)(2.000,19.679){2}{\rule{0.400pt}{3.450pt}}
\put(434.67,231){\rule{0.400pt}{14.936pt}}
\multiput(434.17,262.00)(1.000,-31.000){2}{\rule{0.400pt}{7.468pt}}
\put(436.17,231){\rule{0.400pt}{1.100pt}}
\multiput(435.17,231.00)(2.000,2.717){2}{\rule{0.400pt}{0.550pt}}
\put(437.67,236){\rule{0.400pt}{4.336pt}}
\multiput(437.17,236.00)(1.000,9.000){2}{\rule{0.400pt}{2.168pt}}
\put(439.17,248){\rule{0.400pt}{1.300pt}}
\multiput(438.17,251.30)(2.000,-3.302){2}{\rule{0.400pt}{0.650pt}}
\put(440.67,248){\rule{0.400pt}{1.204pt}}
\multiput(440.17,248.00)(1.000,2.500){2}{\rule{0.400pt}{0.602pt}}
\put(442,251.67){\rule{0.482pt}{0.400pt}}
\multiput(442.00,252.17)(1.000,-1.000){2}{\rule{0.241pt}{0.400pt}}
\put(443.67,252){\rule{0.400pt}{1.686pt}}
\multiput(443.17,252.00)(1.000,3.500){2}{\rule{0.400pt}{0.843pt}}
\put(445.17,253){\rule{0.400pt}{1.300pt}}
\multiput(444.17,256.30)(2.000,-3.302){2}{\rule{0.400pt}{0.650pt}}
\put(446.67,253){\rule{0.400pt}{2.168pt}}
\multiput(446.17,253.00)(1.000,4.500){2}{\rule{0.400pt}{1.084pt}}
\put(448,261.67){\rule{0.482pt}{0.400pt}}
\multiput(448.00,261.17)(1.000,1.000){2}{\rule{0.241pt}{0.400pt}}
\put(449.67,251){\rule{0.400pt}{2.891pt}}
\multiput(449.17,257.00)(1.000,-6.000){2}{\rule{0.400pt}{1.445pt}}
\put(451.17,251){\rule{0.400pt}{1.100pt}}
\multiput(450.17,251.00)(2.000,2.717){2}{\rule{0.400pt}{0.550pt}}
\put(453,254.17){\rule{0.482pt}{0.400pt}}
\multiput(453.00,255.17)(1.000,-2.000){2}{\rule{0.241pt}{0.400pt}}
\put(382.0,270.0){\usebox{\plotpoint}}
\put(456.17,246){\rule{0.400pt}{1.700pt}}
\multiput(455.17,250.47)(2.000,-4.472){2}{\rule{0.400pt}{0.850pt}}
\put(457.67,246){\rule{0.400pt}{1.927pt}}
\multiput(457.17,246.00)(1.000,4.000){2}{\rule{0.400pt}{0.964pt}}
\put(459.17,239){\rule{0.400pt}{3.100pt}}
\multiput(458.17,247.57)(2.000,-8.566){2}{\rule{0.400pt}{1.550pt}}
\put(461,238.67){\rule{0.241pt}{0.400pt}}
\multiput(461.00,238.17)(0.500,1.000){2}{\rule{0.120pt}{0.400pt}}
\put(462.17,240){\rule{0.400pt}{1.300pt}}
\multiput(461.17,240.00)(2.000,3.302){2}{\rule{0.400pt}{0.650pt}}
\put(463.67,244){\rule{0.400pt}{0.482pt}}
\multiput(463.17,245.00)(1.000,-1.000){2}{\rule{0.400pt}{0.241pt}}
\put(465,244.17){\rule{0.482pt}{0.400pt}}
\multiput(465.00,243.17)(1.000,2.000){2}{\rule{0.241pt}{0.400pt}}
\put(455.0,254.0){\usebox{\plotpoint}}
\put(468,245.67){\rule{0.482pt}{0.400pt}}
\multiput(468.00,245.17)(1.000,1.000){2}{\rule{0.241pt}{0.400pt}}
\put(470,245.67){\rule{0.241pt}{0.400pt}}
\multiput(470.00,246.17)(0.500,-1.000){2}{\rule{0.120pt}{0.400pt}}
\put(471,246.17){\rule{0.482pt}{0.400pt}}
\multiput(471.00,245.17)(1.000,2.000){2}{\rule{0.241pt}{0.400pt}}
\put(467.0,246.0){\usebox{\plotpoint}}
\put(474,246.17){\rule{0.482pt}{0.400pt}}
\multiput(474.00,247.17)(1.000,-2.000){2}{\rule{0.241pt}{0.400pt}}
\put(476,245.67){\rule{0.241pt}{0.400pt}}
\multiput(476.00,245.17)(0.500,1.000){2}{\rule{0.120pt}{0.400pt}}
\put(473.0,248.0){\usebox{\plotpoint}}
\put(480,245.67){\rule{0.482pt}{0.400pt}}
\multiput(480.00,246.17)(1.000,-1.000){2}{\rule{0.241pt}{0.400pt}}
\put(482,245.67){\rule{0.241pt}{0.400pt}}
\multiput(482.00,245.17)(0.500,1.000){2}{\rule{0.120pt}{0.400pt}}
\put(483,245.17){\rule{0.482pt}{0.400pt}}
\multiput(483.00,246.17)(1.000,-2.000){2}{\rule{0.241pt}{0.400pt}}
\put(477.0,247.0){\rule[-0.200pt]{0.723pt}{0.400pt}}
\put(486,244.67){\rule{0.482pt}{0.400pt}}
\multiput(486.00,244.17)(1.000,1.000){2}{\rule{0.241pt}{0.400pt}}
\put(485.0,245.0){\usebox{\plotpoint}}
\put(488.0,246.0){\rule[-0.200pt]{3.613pt}{0.400pt}}
\end{picture}

%% file: fig2a3.latex
%%%  IMPORTANT REMARK
%%% 
%%%  This figure is in LaTeX format.
%%%  One must \input it in a LaTeX file and compile it.
%
%
%
% GNUPLOT: LaTeX picture
\setlength{\unitlength}{0.240900pt}
\ifx\plotpoint\undefined\newsavebox{\plotpoint}\fi
\begin{picture}(524,450)(0,0)
\font\gnuplot=cmr10 at 10pt
\gnuplot
\footnotesize
\sbox{\plotpoint}{\rule[-0.200pt]{0.400pt}{0.400pt}}%
\put(180.0,82.0){\rule[-0.200pt]{4.818pt}{0.400pt}}
\put(160,82){\makebox(0,0)[r]{-0.008}}
\put(483.0,82.0){\rule[-0.200pt]{4.818pt}{0.400pt}}
\put(180.0,164.0){\rule[-0.200pt]{4.818pt}{0.400pt}}
\put(160,164){\makebox(0,0)[r]{-0.004}}
\put(483.0,164.0){\rule[-0.200pt]{4.818pt}{0.400pt}}
\put(180.0,246.0){\rule[-0.200pt]{4.818pt}{0.400pt}}
\put(160,246){\makebox(0,0)[r]{0}}
\put(483.0,246.0){\rule[-0.200pt]{4.818pt}{0.400pt}}
\put(180.0,328.0){\rule[-0.200pt]{4.818pt}{0.400pt}}
\put(160,328){\makebox(0,0)[r]{0.004}}
\put(483.0,328.0){\rule[-0.200pt]{4.818pt}{0.400pt}}
\put(180.0,410.0){\rule[-0.200pt]{4.818pt}{0.400pt}}
\put(160,410){\makebox(0,0)[r]{0.008}}
\put(483.0,410.0){\rule[-0.200pt]{4.818pt}{0.400pt}}
\put(180.0,82.0){\rule[-0.200pt]{0.400pt}{4.818pt}}
\put(180,41){\makebox(0,0){900}}
\put(180.0,390.0){\rule[-0.200pt]{0.400pt}{4.818pt}}
\put(341.0,82.0){\rule[-0.200pt]{0.400pt}{4.818pt}}
\put(341,41){\makebox(0,0){1300}}
\put(341.0,390.0){\rule[-0.200pt]{0.400pt}{4.818pt}}
\put(503.0,82.0){\rule[-0.200pt]{0.400pt}{4.818pt}}
\put(503,41){\makebox(0,0){1700}}
\put(503.0,390.0){\rule[-0.200pt]{0.400pt}{4.818pt}}
\put(180.0,82.0){\rule[-0.200pt]{77.811pt}{0.400pt}}
\put(503.0,82.0){\rule[-0.200pt]{0.400pt}{79.015pt}}
\put(180.0,410.0){\rule[-0.200pt]{77.811pt}{0.400pt}}
\put(180.0,82.0){\rule[-0.200pt]{0.400pt}{79.015pt}}
\put(180,246){\usebox{\plotpoint}}
\put(180,246){\usebox{\plotpoint}}
\put(180.0,246.0){\rule[-0.200pt]{4.577pt}{0.400pt}}
\put(199.0,246.0){\usebox{\plotpoint}}
\put(199.0,247.0){\usebox{\plotpoint}}
\put(200.0,246.0){\usebox{\plotpoint}}
\put(206,244.67){\rule{0.241pt}{0.400pt}}
\multiput(206.00,245.17)(0.500,-1.000){2}{\rule{0.120pt}{0.400pt}}
\put(200.0,246.0){\rule[-0.200pt]{1.445pt}{0.400pt}}
\put(207.0,245.0){\usebox{\plotpoint}}
\put(207.0,245.0){\usebox{\plotpoint}}
\put(207.0,245.0){\rule[-0.200pt]{0.482pt}{0.400pt}}
\put(209,244.67){\rule{0.241pt}{0.400pt}}
\multiput(209.00,245.17)(0.500,-1.000){2}{\rule{0.120pt}{0.400pt}}
\put(209.0,245.0){\usebox{\plotpoint}}
\put(210,245){\usebox{\plotpoint}}
\put(210.0,245.0){\rule[-0.200pt]{0.482pt}{0.400pt}}
\put(212.0,244.0){\usebox{\plotpoint}}
\put(212.0,244.0){\usebox{\plotpoint}}
\put(213.0,244.0){\usebox{\plotpoint}}
\put(214,244.67){\rule{0.241pt}{0.400pt}}
\multiput(214.00,244.17)(0.500,1.000){2}{\rule{0.120pt}{0.400pt}}
\put(213.0,245.0){\usebox{\plotpoint}}
\put(215,244.67){\rule{0.241pt}{0.400pt}}
\multiput(215.00,244.17)(0.500,1.000){2}{\rule{0.120pt}{0.400pt}}
\put(215.0,245.0){\usebox{\plotpoint}}
\put(216,246){\usebox{\plotpoint}}
\put(216,246){\usebox{\plotpoint}}
\put(219,245.67){\rule{0.241pt}{0.400pt}}
\multiput(219.00,245.17)(0.500,1.000){2}{\rule{0.120pt}{0.400pt}}
\put(216.0,246.0){\rule[-0.200pt]{0.723pt}{0.400pt}}
\put(220.0,246.0){\usebox{\plotpoint}}
\put(220.0,246.0){\usebox{\plotpoint}}
\put(220.0,247.0){\rule[-0.200pt]{0.482pt}{0.400pt}}
\put(221.67,246){\rule{0.400pt}{0.482pt}}
\multiput(221.17,246.00)(1.000,1.000){2}{\rule{0.400pt}{0.241pt}}
\put(222.0,246.0){\usebox{\plotpoint}}
\put(223,248){\usebox{\plotpoint}}
\put(223.0,248.0){\usebox{\plotpoint}}
\put(223.67,248){\rule{0.400pt}{0.482pt}}
\multiput(223.17,249.00)(1.000,-1.000){2}{\rule{0.400pt}{0.241pt}}
\put(224.0,248.0){\rule[-0.200pt]{0.400pt}{0.482pt}}
\put(225.0,248.0){\rule[-0.200pt]{0.400pt}{1.204pt}}
\put(225.0,253.0){\usebox{\plotpoint}}
\put(226.0,245.0){\rule[-0.200pt]{0.400pt}{1.927pt}}
\put(225.67,245){\rule{0.400pt}{0.964pt}}
\multiput(225.17,247.00)(1.000,-2.000){2}{\rule{0.400pt}{0.482pt}}
\put(226.0,245.0){\rule[-0.200pt]{0.400pt}{0.964pt}}
\put(227,245){\usebox{\plotpoint}}
\put(227.0,245.0){\usebox{\plotpoint}}
\put(228.0,245.0){\usebox{\plotpoint}}
\put(228,241.67){\rule{0.241pt}{0.400pt}}
\multiput(228.00,241.17)(0.500,1.000){2}{\rule{0.120pt}{0.400pt}}
\put(228.0,242.0){\rule[-0.200pt]{0.400pt}{0.964pt}}
\put(229.0,242.0){\usebox{\plotpoint}}
\put(229.0,242.0){\usebox{\plotpoint}}
\put(229.67,240){\rule{0.400pt}{0.964pt}}
\multiput(229.17,242.00)(1.000,-2.000){2}{\rule{0.400pt}{0.482pt}}
\put(230.0,242.0){\rule[-0.200pt]{0.400pt}{0.482pt}}
\put(230.67,239){\rule{0.400pt}{0.964pt}}
\multiput(230.17,241.00)(1.000,-2.000){2}{\rule{0.400pt}{0.482pt}}
\put(231.0,240.0){\rule[-0.200pt]{0.400pt}{0.723pt}}
\put(232,239){\usebox{\plotpoint}}
\put(232,239){\usebox{\plotpoint}}
\put(232,238.67){\rule{0.241pt}{0.400pt}}
\multiput(232.00,238.17)(0.500,1.000){2}{\rule{0.120pt}{0.400pt}}
\put(233.0,238.0){\rule[-0.200pt]{0.400pt}{0.482pt}}
\put(233.0,238.0){\usebox{\plotpoint}}
\put(234,238.67){\rule{0.241pt}{0.400pt}}
\multiput(234.00,239.17)(0.500,-1.000){2}{\rule{0.120pt}{0.400pt}}
\put(234.0,238.0){\rule[-0.200pt]{0.400pt}{0.482pt}}
\put(234.67,236){\rule{0.400pt}{1.204pt}}
\multiput(234.17,238.50)(1.000,-2.500){2}{\rule{0.400pt}{0.602pt}}
\put(235.0,239.0){\rule[-0.200pt]{0.400pt}{0.482pt}}
\put(236,237.67){\rule{0.241pt}{0.400pt}}
\multiput(236.00,238.17)(0.500,-1.000){2}{\rule{0.120pt}{0.400pt}}
\put(236.0,236.0){\rule[-0.200pt]{0.400pt}{0.723pt}}
\put(236.67,234){\rule{0.400pt}{0.964pt}}
\multiput(236.17,234.00)(1.000,2.000){2}{\rule{0.400pt}{0.482pt}}
\put(237.0,234.0){\rule[-0.200pt]{0.400pt}{0.964pt}}
\put(238,229.67){\rule{0.241pt}{0.400pt}}
\multiput(238.00,229.17)(0.500,1.000){2}{\rule{0.120pt}{0.400pt}}
\put(238.0,230.0){\rule[-0.200pt]{0.400pt}{1.927pt}}
\put(239.0,231.0){\rule[-0.200pt]{0.400pt}{1.445pt}}
\put(238.67,234){\rule{0.400pt}{0.723pt}}
\multiput(238.17,234.00)(1.000,1.500){2}{\rule{0.400pt}{0.361pt}}
\put(239.0,234.0){\rule[-0.200pt]{0.400pt}{0.723pt}}
\put(240,237){\usebox{\plotpoint}}
\put(240.0,237.0){\usebox{\plotpoint}}
\put(241.0,236.0){\usebox{\plotpoint}}
\put(241,237.67){\rule{0.241pt}{0.400pt}}
\multiput(241.00,238.17)(0.500,-1.000){2}{\rule{0.120pt}{0.400pt}}
\put(241.0,236.0){\rule[-0.200pt]{0.400pt}{0.723pt}}
\put(242.0,238.0){\usebox{\plotpoint}}
\put(242.0,239.0){\usebox{\plotpoint}}
\put(242.67,238){\rule{0.400pt}{0.482pt}}
\multiput(242.17,238.00)(1.000,1.000){2}{\rule{0.400pt}{0.241pt}}
\put(243.0,238.0){\usebox{\plotpoint}}
\put(243.67,239){\rule{0.400pt}{0.482pt}}
\multiput(243.17,239.00)(1.000,1.000){2}{\rule{0.400pt}{0.241pt}}
\put(244.0,239.0){\usebox{\plotpoint}}
\put(245,241){\usebox{\plotpoint}}
\put(244.67,241){\rule{0.400pt}{0.482pt}}
\multiput(244.17,242.00)(1.000,-1.000){2}{\rule{0.400pt}{0.241pt}}
\put(245.0,241.0){\rule[-0.200pt]{0.400pt}{0.482pt}}
\put(246.0,241.0){\rule[-0.200pt]{0.400pt}{0.482pt}}
\put(246.0,243.0){\usebox{\plotpoint}}
\put(247.0,241.0){\rule[-0.200pt]{0.400pt}{0.482pt}}
\put(247,240.67){\rule{0.241pt}{0.400pt}}
\multiput(247.00,241.17)(0.500,-1.000){2}{\rule{0.120pt}{0.400pt}}
\put(247.0,241.0){\usebox{\plotpoint}}
\put(247.67,240){\rule{0.400pt}{1.927pt}}
\multiput(247.17,240.00)(1.000,4.000){2}{\rule{0.400pt}{0.964pt}}
\put(248.0,240.0){\usebox{\plotpoint}}
\put(249.0,244.0){\rule[-0.200pt]{0.400pt}{0.964pt}}
\put(249.0,244.0){\usebox{\plotpoint}}
\put(249.0,245.0){\usebox{\plotpoint}}
\put(249.67,246){\rule{0.400pt}{1.927pt}}
\multiput(249.17,250.00)(1.000,-4.000){2}{\rule{0.400pt}{0.964pt}}
\put(250.0,245.0){\rule[-0.200pt]{0.400pt}{2.168pt}}
\put(250.67,232){\rule{0.400pt}{7.950pt}}
\multiput(250.17,248.50)(1.000,-16.500){2}{\rule{0.400pt}{3.975pt}}
\put(251.0,246.0){\rule[-0.200pt]{0.400pt}{4.577pt}}
\put(251.67,231){\rule{0.400pt}{4.336pt}}
\multiput(251.17,240.00)(1.000,-9.000){2}{\rule{0.400pt}{2.168pt}}
\put(252.0,232.0){\rule[-0.200pt]{0.400pt}{4.095pt}}
\put(252.67,230){\rule{0.400pt}{1.927pt}}
\multiput(252.17,230.00)(1.000,4.000){2}{\rule{0.400pt}{0.964pt}}
\put(253.0,230.0){\usebox{\plotpoint}}
\put(253.67,218){\rule{0.400pt}{0.964pt}}
\multiput(253.17,218.00)(1.000,2.000){2}{\rule{0.400pt}{0.482pt}}
\put(254.0,218.0){\rule[-0.200pt]{0.400pt}{4.818pt}}
\put(255,217.67){\rule{0.241pt}{0.400pt}}
\multiput(255.00,218.17)(0.500,-1.000){2}{\rule{0.120pt}{0.400pt}}
\put(255.0,219.0){\rule[-0.200pt]{0.400pt}{0.723pt}}
\put(256.0,218.0){\rule[-0.200pt]{0.400pt}{1.927pt}}
\put(255.67,210){\rule{0.400pt}{3.614pt}}
\multiput(255.17,217.50)(1.000,-7.500){2}{\rule{0.400pt}{1.807pt}}
\put(256.0,225.0){\usebox{\plotpoint}}
\put(256.67,201){\rule{0.400pt}{5.059pt}}
\multiput(256.17,211.50)(1.000,-10.500){2}{\rule{0.400pt}{2.529pt}}
\put(257.0,210.0){\rule[-0.200pt]{0.400pt}{2.891pt}}
\put(258,207.67){\rule{0.241pt}{0.400pt}}
\multiput(258.00,208.17)(0.500,-1.000){2}{\rule{0.120pt}{0.400pt}}
\put(258.0,201.0){\rule[-0.200pt]{0.400pt}{1.927pt}}
\put(259,205.67){\rule{0.241pt}{0.400pt}}
\multiput(259.00,206.17)(0.500,-1.000){2}{\rule{0.120pt}{0.400pt}}
\put(259.0,207.0){\usebox{\plotpoint}}
\put(260.0,206.0){\rule[-0.200pt]{0.400pt}{1.927pt}}
\put(259.67,208){\rule{0.400pt}{2.409pt}}
\multiput(259.17,208.00)(1.000,5.000){2}{\rule{0.400pt}{1.204pt}}
\put(260.0,208.0){\rule[-0.200pt]{0.400pt}{1.445pt}}
\put(260.67,205){\rule{0.400pt}{3.373pt}}
\multiput(260.17,212.00)(1.000,-7.000){2}{\rule{0.400pt}{1.686pt}}
\put(261.0,218.0){\usebox{\plotpoint}}
\put(262.0,205.0){\rule[-0.200pt]{0.400pt}{1.686pt}}
\put(262.0,207.0){\rule[-0.200pt]{0.400pt}{1.204pt}}
\put(262.0,207.0){\usebox{\plotpoint}}
\put(262.67,203){\rule{0.400pt}{1.204pt}}
\multiput(262.17,203.00)(1.000,2.500){2}{\rule{0.400pt}{0.602pt}}
\put(263.0,203.0){\rule[-0.200pt]{0.400pt}{0.964pt}}
\put(264.0,199.0){\rule[-0.200pt]{0.400pt}{2.168pt}}
\put(263.67,191){\rule{0.400pt}{2.650pt}}
\multiput(263.17,196.50)(1.000,-5.500){2}{\rule{0.400pt}{1.325pt}}
\put(264.0,199.0){\rule[-0.200pt]{0.400pt}{0.723pt}}
\put(264.67,193){\rule{0.400pt}{0.964pt}}
\multiput(264.17,195.00)(1.000,-2.000){2}{\rule{0.400pt}{0.482pt}}
\put(265.0,191.0){\rule[-0.200pt]{0.400pt}{1.445pt}}
\put(265.67,188){\rule{0.400pt}{1.445pt}}
\multiput(265.17,188.00)(1.000,3.000){2}{\rule{0.400pt}{0.723pt}}
\put(266.0,188.0){\rule[-0.200pt]{0.400pt}{1.204pt}}
\put(267,178.67){\rule{0.241pt}{0.400pt}}
\multiput(267.00,178.17)(0.500,1.000){2}{\rule{0.120pt}{0.400pt}}
\put(267.0,179.0){\rule[-0.200pt]{0.400pt}{3.613pt}}
\put(268.0,180.0){\rule[-0.200pt]{0.400pt}{2.891pt}}
\put(267.67,185){\rule{0.400pt}{1.927pt}}
\multiput(267.17,185.00)(1.000,4.000){2}{\rule{0.400pt}{0.964pt}}
\put(268.0,185.0){\rule[-0.200pt]{0.400pt}{1.686pt}}
\put(269,193){\usebox{\plotpoint}}
\put(269.0,193.0){\usebox{\plotpoint}}
\put(270.0,190.0){\rule[-0.200pt]{0.400pt}{0.723pt}}
\put(269.67,200){\rule{0.400pt}{0.482pt}}
\multiput(269.17,200.00)(1.000,1.000){2}{\rule{0.400pt}{0.241pt}}
\put(270.0,190.0){\rule[-0.200pt]{0.400pt}{2.409pt}}
\put(270.67,174){\rule{0.400pt}{3.854pt}}
\multiput(270.17,174.00)(1.000,8.000){2}{\rule{0.400pt}{1.927pt}}
\put(271.0,174.0){\rule[-0.200pt]{0.400pt}{6.745pt}}
\put(271.67,169){\rule{0.400pt}{2.409pt}}
\multiput(271.17,169.00)(1.000,5.000){2}{\rule{0.400pt}{1.204pt}}
\put(272.0,169.0){\rule[-0.200pt]{0.400pt}{5.059pt}}
\put(273,179){\usebox{\plotpoint}}
\put(272.67,173){\rule{0.400pt}{1.445pt}}
\multiput(272.17,176.00)(1.000,-3.000){2}{\rule{0.400pt}{0.723pt}}
\put(274.0,173.0){\rule[-0.200pt]{0.400pt}{1.686pt}}
\put(273.67,151){\rule{0.400pt}{2.650pt}}
\multiput(273.17,151.00)(1.000,5.500){2}{\rule{0.400pt}{1.325pt}}
\put(274.0,151.0){\rule[-0.200pt]{0.400pt}{6.986pt}}
\put(275,162){\usebox{\plotpoint}}
\put(275,160.67){\rule{0.241pt}{0.400pt}}
\multiput(275.00,161.17)(0.500,-1.000){2}{\rule{0.120pt}{0.400pt}}
\put(276.0,132.0){\rule[-0.200pt]{0.400pt}{6.986pt}}
\put(275.67,104){\rule{0.400pt}{13.490pt}}
\multiput(275.17,132.00)(1.000,-28.000){2}{\rule{0.400pt}{6.745pt}}
\put(276.0,132.0){\rule[-0.200pt]{0.400pt}{6.745pt}}
\put(277,104){\usebox{\plotpoint}}
\put(276.67,104){\rule{0.400pt}{13.731pt}}
\multiput(276.17,104.00)(1.000,28.500){2}{\rule{0.400pt}{6.866pt}}
\put(277.67,133){\rule{0.400pt}{6.745pt}}
\multiput(277.17,133.00)(1.000,14.000){2}{\rule{0.400pt}{3.373pt}}
\put(278.0,133.0){\rule[-0.200pt]{0.400pt}{6.745pt}}
\put(278.67,152){\rule{0.400pt}{3.132pt}}
\multiput(278.17,158.50)(1.000,-6.500){2}{\rule{0.400pt}{1.566pt}}
\put(279.0,161.0){\rule[-0.200pt]{0.400pt}{0.964pt}}
\put(279.67,178){\rule{0.400pt}{1.445pt}}
\multiput(279.17,181.00)(1.000,-3.000){2}{\rule{0.400pt}{0.723pt}}
\put(280.0,152.0){\rule[-0.200pt]{0.400pt}{7.709pt}}
\put(281,178){\usebox{\plotpoint}}
\put(280.67,169){\rule{0.400pt}{2.891pt}}
\multiput(280.17,175.00)(1.000,-6.000){2}{\rule{0.400pt}{1.445pt}}
\put(281.0,178.0){\rule[-0.200pt]{0.400pt}{0.723pt}}
\put(281.67,170){\rule{0.400pt}{4.336pt}}
\multiput(281.17,170.00)(1.000,9.000){2}{\rule{0.400pt}{2.168pt}}
\put(282.0,169.0){\usebox{\plotpoint}}
\put(283.0,174.0){\rule[-0.200pt]{0.400pt}{3.373pt}}
\put(282.67,193){\rule{0.400pt}{0.482pt}}
\multiput(282.17,194.00)(1.000,-1.000){2}{\rule{0.400pt}{0.241pt}}
\put(283.0,174.0){\rule[-0.200pt]{0.400pt}{5.059pt}}
\put(283.67,192){\rule{0.400pt}{0.964pt}}
\multiput(283.17,194.00)(1.000,-2.000){2}{\rule{0.400pt}{0.482pt}}
\put(284.0,193.0){\rule[-0.200pt]{0.400pt}{0.723pt}}
\put(284.67,190){\rule{0.400pt}{2.650pt}}
\multiput(284.17,195.50)(1.000,-5.500){2}{\rule{0.400pt}{1.325pt}}
\put(285.0,192.0){\rule[-0.200pt]{0.400pt}{2.168pt}}
\put(285.67,188){\rule{0.400pt}{1.686pt}}
\multiput(285.17,191.50)(1.000,-3.500){2}{\rule{0.400pt}{0.843pt}}
\put(286.0,190.0){\rule[-0.200pt]{0.400pt}{1.204pt}}
\put(287.0,185.0){\rule[-0.200pt]{0.400pt}{0.723pt}}
\put(286.67,199){\rule{0.400pt}{2.409pt}}
\multiput(286.17,204.00)(1.000,-5.000){2}{\rule{0.400pt}{1.204pt}}
\put(287.0,185.0){\rule[-0.200pt]{0.400pt}{5.782pt}}
\put(288,202.67){\rule{0.241pt}{0.400pt}}
\multiput(288.00,202.17)(0.500,1.000){2}{\rule{0.120pt}{0.400pt}}
\put(288.0,199.0){\rule[-0.200pt]{0.400pt}{0.964pt}}
\put(289.0,204.0){\rule[-0.200pt]{0.400pt}{3.854pt}}
\put(288.67,203){\rule{0.400pt}{8.191pt}}
\multiput(288.17,203.00)(1.000,17.000){2}{\rule{0.400pt}{4.095pt}}
\put(289.0,203.0){\rule[-0.200pt]{0.400pt}{4.095pt}}
\put(289.67,216){\rule{0.400pt}{4.336pt}}
\multiput(289.17,225.00)(1.000,-9.000){2}{\rule{0.400pt}{2.168pt}}
\put(290.0,234.0){\rule[-0.200pt]{0.400pt}{0.723pt}}
\put(291.0,216.0){\rule[-0.200pt]{0.400pt}{2.168pt}}
\put(291.0,214.0){\rule[-0.200pt]{0.400pt}{2.650pt}}
\put(291.0,214.0){\usebox{\plotpoint}}
\put(292,216.67){\rule{0.241pt}{0.400pt}}
\multiput(292.00,216.17)(0.500,1.000){2}{\rule{0.120pt}{0.400pt}}
\put(292.0,214.0){\rule[-0.200pt]{0.400pt}{0.723pt}}
\put(293.0,214.0){\rule[-0.200pt]{0.400pt}{0.964pt}}
\put(292.67,208){\rule{0.400pt}{1.927pt}}
\multiput(292.17,212.00)(1.000,-4.000){2}{\rule{0.400pt}{0.964pt}}
\put(293.0,214.0){\rule[-0.200pt]{0.400pt}{0.482pt}}
\put(294.0,208.0){\rule[-0.200pt]{0.400pt}{0.723pt}}
\put(294.0,211.0){\usebox{\plotpoint}}
\put(294.67,203){\rule{0.400pt}{1.686pt}}
\multiput(294.17,203.00)(1.000,3.500){2}{\rule{0.400pt}{0.843pt}}
\put(295.0,203.0){\rule[-0.200pt]{0.400pt}{1.927pt}}
\put(296.0,195.0){\rule[-0.200pt]{0.400pt}{3.613pt}}
\put(296.0,195.0){\usebox{\plotpoint}}
\put(297.0,195.0){\rule[-0.200pt]{0.400pt}{3.613pt}}
\put(296.67,202){\rule{0.400pt}{2.168pt}}
\multiput(296.17,202.00)(1.000,4.500){2}{\rule{0.400pt}{1.084pt}}
\put(297.0,202.0){\rule[-0.200pt]{0.400pt}{1.927pt}}
\put(298,211){\usebox{\plotpoint}}
\put(298,209.67){\rule{0.241pt}{0.400pt}}
\multiput(298.00,210.17)(0.500,-1.000){2}{\rule{0.120pt}{0.400pt}}
\put(298.67,208){\rule{0.400pt}{1.445pt}}
\multiput(298.17,208.00)(1.000,3.000){2}{\rule{0.400pt}{0.723pt}}
\put(299.0,208.0){\rule[-0.200pt]{0.400pt}{0.482pt}}
\put(300.0,212.0){\rule[-0.200pt]{0.400pt}{0.482pt}}
\put(300,214.67){\rule{0.241pt}{0.400pt}}
\multiput(300.00,215.17)(0.500,-1.000){2}{\rule{0.120pt}{0.400pt}}
\put(300.0,212.0){\rule[-0.200pt]{0.400pt}{0.964pt}}
\put(301,212.67){\rule{0.241pt}{0.400pt}}
\multiput(301.00,212.17)(0.500,1.000){2}{\rule{0.120pt}{0.400pt}}
\put(301.0,213.0){\rule[-0.200pt]{0.400pt}{0.482pt}}
\put(302.0,214.0){\rule[-0.200pt]{0.400pt}{1.445pt}}
\put(301.67,215){\rule{0.400pt}{2.409pt}}
\multiput(301.17,215.00)(1.000,5.000){2}{\rule{0.400pt}{1.204pt}}
\put(302.0,215.0){\rule[-0.200pt]{0.400pt}{1.204pt}}
\put(302.67,221){\rule{0.400pt}{0.723pt}}
\multiput(302.17,222.50)(1.000,-1.500){2}{\rule{0.400pt}{0.361pt}}
\put(303.0,224.0){\usebox{\plotpoint}}
\put(304.0,221.0){\usebox{\plotpoint}}
\put(304,220.67){\rule{0.241pt}{0.400pt}}
\multiput(304.00,220.17)(0.500,1.000){2}{\rule{0.120pt}{0.400pt}}
\put(304.0,221.0){\usebox{\plotpoint}}
\put(304.67,219){\rule{0.400pt}{0.482pt}}
\multiput(304.17,219.00)(1.000,1.000){2}{\rule{0.400pt}{0.241pt}}
\put(305.0,219.0){\rule[-0.200pt]{0.400pt}{0.723pt}}
\put(305.67,218){\rule{0.400pt}{1.204pt}}
\multiput(305.17,218.00)(1.000,2.500){2}{\rule{0.400pt}{0.602pt}}
\put(306.0,218.0){\rule[-0.200pt]{0.400pt}{0.723pt}}
\put(307,220.67){\rule{0.241pt}{0.400pt}}
\multiput(307.00,220.17)(0.500,1.000){2}{\rule{0.120pt}{0.400pt}}
\put(307.0,221.0){\rule[-0.200pt]{0.400pt}{0.482pt}}
\put(308,222){\usebox{\plotpoint}}
\put(307.67,222){\rule{0.400pt}{0.964pt}}
\multiput(307.17,224.00)(1.000,-2.000){2}{\rule{0.400pt}{0.482pt}}
\put(308.0,222.0){\rule[-0.200pt]{0.400pt}{0.964pt}}
\put(309,229.67){\rule{0.241pt}{0.400pt}}
\multiput(309.00,229.17)(0.500,1.000){2}{\rule{0.120pt}{0.400pt}}
\put(309.0,222.0){\rule[-0.200pt]{0.400pt}{1.927pt}}
\put(310.0,218.0){\rule[-0.200pt]{0.400pt}{3.132pt}}
\put(309.67,218){\rule{0.400pt}{1.686pt}}
\multiput(309.17,221.50)(1.000,-3.500){2}{\rule{0.400pt}{0.843pt}}
\put(310.0,218.0){\rule[-0.200pt]{0.400pt}{1.686pt}}
\put(311,216.67){\rule{0.241pt}{0.400pt}}
\multiput(311.00,216.17)(0.500,1.000){2}{\rule{0.120pt}{0.400pt}}
\put(311.0,217.0){\usebox{\plotpoint}}
\put(312.0,218.0){\rule[-0.200pt]{0.400pt}{0.482pt}}
\put(311.67,215){\rule{0.400pt}{0.482pt}}
\multiput(311.17,215.00)(1.000,1.000){2}{\rule{0.400pt}{0.241pt}}
\put(312.0,215.0){\rule[-0.200pt]{0.400pt}{1.204pt}}
\put(312.67,211){\rule{0.400pt}{0.482pt}}
\multiput(312.17,211.00)(1.000,1.000){2}{\rule{0.400pt}{0.241pt}}
\put(313.0,211.0){\rule[-0.200pt]{0.400pt}{1.445pt}}
\put(313.67,204){\rule{0.400pt}{2.409pt}}
\multiput(313.17,209.00)(1.000,-5.000){2}{\rule{0.400pt}{1.204pt}}
\put(314.0,213.0){\usebox{\plotpoint}}
\put(314.67,195){\rule{0.400pt}{4.336pt}}
\multiput(314.17,204.00)(1.000,-9.000){2}{\rule{0.400pt}{2.168pt}}
\put(315.0,204.0){\rule[-0.200pt]{0.400pt}{2.168pt}}
\put(316,195){\usebox{\plotpoint}}
\put(315.67,206){\rule{0.400pt}{2.891pt}}
\multiput(315.17,212.00)(1.000,-6.000){2}{\rule{0.400pt}{1.445pt}}
\put(316.0,195.0){\rule[-0.200pt]{0.400pt}{5.541pt}}
\put(317.0,206.0){\rule[-0.200pt]{0.400pt}{2.891pt}}
\put(317.0,218.0){\usebox{\plotpoint}}
\put(318.0,218.0){\rule[-0.200pt]{0.400pt}{0.723pt}}
\put(317.67,214){\rule{0.400pt}{3.854pt}}
\multiput(317.17,214.00)(1.000,8.000){2}{\rule{0.400pt}{1.927pt}}
\put(318.0,214.0){\rule[-0.200pt]{0.400pt}{1.686pt}}
\put(318.67,224){\rule{0.400pt}{0.723pt}}
\multiput(318.17,225.50)(1.000,-1.500){2}{\rule{0.400pt}{0.361pt}}
\put(319.0,227.0){\rule[-0.200pt]{0.400pt}{0.723pt}}
\put(319.67,220){\rule{0.400pt}{1.686pt}}
\multiput(319.17,223.50)(1.000,-3.500){2}{\rule{0.400pt}{0.843pt}}
\put(320.0,224.0){\rule[-0.200pt]{0.400pt}{0.723pt}}
\put(320.67,223){\rule{0.400pt}{0.964pt}}
\multiput(320.17,225.00)(1.000,-2.000){2}{\rule{0.400pt}{0.482pt}}
\put(321.0,220.0){\rule[-0.200pt]{0.400pt}{1.686pt}}
\put(322,227.67){\rule{0.241pt}{0.400pt}}
\multiput(322.00,228.17)(0.500,-1.000){2}{\rule{0.120pt}{0.400pt}}
\put(322.0,223.0){\rule[-0.200pt]{0.400pt}{1.445pt}}
\put(323.0,228.0){\rule[-0.200pt]{0.400pt}{1.204pt}}
\put(322.67,229){\rule{0.400pt}{1.445pt}}
\multiput(322.17,229.00)(1.000,3.000){2}{\rule{0.400pt}{0.723pt}}
\put(323.0,229.0){\rule[-0.200pt]{0.400pt}{0.964pt}}
\put(323.67,230){\rule{0.400pt}{1.445pt}}
\multiput(323.17,233.00)(1.000,-3.000){2}{\rule{0.400pt}{0.723pt}}
\put(324.0,235.0){\usebox{\plotpoint}}
\put(325.0,230.0){\rule[-0.200pt]{0.400pt}{0.482pt}}
\put(324.67,229){\rule{0.400pt}{0.482pt}}
\multiput(324.17,230.00)(1.000,-1.000){2}{\rule{0.400pt}{0.241pt}}
\put(325.0,231.0){\usebox{\plotpoint}}
\put(325.67,235){\rule{0.400pt}{1.204pt}}
\multiput(325.17,237.50)(1.000,-2.500){2}{\rule{0.400pt}{0.602pt}}
\put(326.0,229.0){\rule[-0.200pt]{0.400pt}{2.650pt}}
\put(326.67,237){\rule{0.400pt}{1.445pt}}
\multiput(326.17,237.00)(1.000,3.000){2}{\rule{0.400pt}{0.723pt}}
\put(327.0,235.0){\rule[-0.200pt]{0.400pt}{0.482pt}}
\put(327.67,237){\rule{0.400pt}{2.650pt}}
\multiput(327.17,237.00)(1.000,5.500){2}{\rule{0.400pt}{1.325pt}}
\put(328.0,237.0){\rule[-0.200pt]{0.400pt}{1.445pt}}
\put(329.0,244.0){\rule[-0.200pt]{0.400pt}{0.964pt}}
\put(328.67,249){\rule{0.400pt}{1.445pt}}
\multiput(328.17,252.00)(1.000,-3.000){2}{\rule{0.400pt}{0.723pt}}
\put(329.0,244.0){\rule[-0.200pt]{0.400pt}{2.650pt}}
\put(330,254.67){\rule{0.241pt}{0.400pt}}
\multiput(330.00,254.17)(0.500,1.000){2}{\rule{0.120pt}{0.400pt}}
\put(330.0,249.0){\rule[-0.200pt]{0.400pt}{1.445pt}}
\put(331.0,256.0){\usebox{\plotpoint}}
\put(330.67,252){\rule{0.400pt}{2.650pt}}
\multiput(330.17,252.00)(1.000,5.500){2}{\rule{0.400pt}{1.325pt}}
\put(331.0,252.0){\rule[-0.200pt]{0.400pt}{1.204pt}}
\put(332,259.67){\rule{0.241pt}{0.400pt}}
\multiput(332.00,260.17)(0.500,-1.000){2}{\rule{0.120pt}{0.400pt}}
\put(332.0,261.0){\rule[-0.200pt]{0.400pt}{0.482pt}}
\put(333.0,260.0){\rule[-0.200pt]{0.400pt}{0.482pt}}
\put(333,255.67){\rule{0.241pt}{0.400pt}}
\multiput(333.00,255.17)(0.500,1.000){2}{\rule{0.120pt}{0.400pt}}
\put(333.0,256.0){\rule[-0.200pt]{0.400pt}{1.445pt}}
\put(333.67,259){\rule{0.400pt}{0.964pt}}
\multiput(333.17,261.00)(1.000,-2.000){2}{\rule{0.400pt}{0.482pt}}
\put(334.0,257.0){\rule[-0.200pt]{0.400pt}{1.445pt}}
\put(335.0,259.0){\rule[-0.200pt]{0.400pt}{1.204pt}}
\put(334.67,263){\rule{0.400pt}{1.445pt}}
\multiput(334.17,263.00)(1.000,3.000){2}{\rule{0.400pt}{0.723pt}}
\put(335.0,263.0){\usebox{\plotpoint}}
\put(335.67,265){\rule{0.400pt}{1.445pt}}
\multiput(335.17,265.00)(1.000,3.000){2}{\rule{0.400pt}{0.723pt}}
\put(336.0,265.0){\rule[-0.200pt]{0.400pt}{0.964pt}}
\put(337.0,271.0){\usebox{\plotpoint}}
\put(336.67,265){\rule{0.400pt}{0.723pt}}
\multiput(336.17,265.00)(1.000,1.500){2}{\rule{0.400pt}{0.361pt}}
\put(337.0,265.0){\rule[-0.200pt]{0.400pt}{1.686pt}}
\put(337.67,262){\rule{0.400pt}{0.723pt}}
\multiput(337.17,263.50)(1.000,-1.500){2}{\rule{0.400pt}{0.361pt}}
\put(338.0,265.0){\rule[-0.200pt]{0.400pt}{0.723pt}}
\put(339.0,262.0){\rule[-0.200pt]{0.400pt}{3.854pt}}
\put(338.67,271){\rule{0.400pt}{0.723pt}}
\multiput(338.17,271.00)(1.000,1.500){2}{\rule{0.400pt}{0.361pt}}
\put(339.0,271.0){\rule[-0.200pt]{0.400pt}{1.686pt}}
\put(340,274){\usebox{\plotpoint}}
\put(339.67,274){\rule{0.400pt}{2.891pt}}
\multiput(339.17,274.00)(1.000,6.000){2}{\rule{0.400pt}{1.445pt}}
\put(341.0,274.0){\rule[-0.200pt]{0.400pt}{2.891pt}}
\put(341.0,274.0){\rule[-0.200pt]{0.400pt}{5.541pt}}
\put(341.0,297.0){\usebox{\plotpoint}}
\put(341.67,279){\rule{0.400pt}{2.168pt}}
\multiput(341.17,279.00)(1.000,4.500){2}{\rule{0.400pt}{1.084pt}}
\put(342.0,279.0){\rule[-0.200pt]{0.400pt}{4.336pt}}
\put(343.0,278.0){\rule[-0.200pt]{0.400pt}{2.409pt}}
\put(343.0,278.0){\usebox{\plotpoint}}
\put(343.67,275){\rule{0.400pt}{1.445pt}}
\multiput(343.17,278.00)(1.000,-3.000){2}{\rule{0.400pt}{0.723pt}}
\put(344.0,278.0){\rule[-0.200pt]{0.400pt}{0.723pt}}
\put(344.67,272){\rule{0.400pt}{1.204pt}}
\multiput(344.17,274.50)(1.000,-2.500){2}{\rule{0.400pt}{0.602pt}}
\put(345.0,275.0){\rule[-0.200pt]{0.400pt}{0.482pt}}
\put(346,273.67){\rule{0.241pt}{0.400pt}}
\multiput(346.00,274.17)(0.500,-1.000){2}{\rule{0.120pt}{0.400pt}}
\put(346.0,272.0){\rule[-0.200pt]{0.400pt}{0.723pt}}
\put(346.67,267){\rule{0.400pt}{1.686pt}}
\multiput(346.17,267.00)(1.000,3.500){2}{\rule{0.400pt}{0.843pt}}
\put(347.0,267.0){\rule[-0.200pt]{0.400pt}{1.686pt}}
\put(348.0,261.0){\rule[-0.200pt]{0.400pt}{3.132pt}}
\put(347.67,262){\rule{0.400pt}{1.927pt}}
\multiput(347.17,262.00)(1.000,4.000){2}{\rule{0.400pt}{0.964pt}}
\put(348.0,261.0){\usebox{\plotpoint}}
\put(348.67,266){\rule{0.400pt}{0.964pt}}
\multiput(348.17,266.00)(1.000,2.000){2}{\rule{0.400pt}{0.482pt}}
\put(349.0,266.0){\rule[-0.200pt]{0.400pt}{0.964pt}}
\put(350,270){\usebox{\plotpoint}}
\put(349.67,269){\rule{0.400pt}{0.482pt}}
\multiput(349.17,270.00)(1.000,-1.000){2}{\rule{0.400pt}{0.241pt}}
\put(350.0,270.0){\usebox{\plotpoint}}
\put(351.0,269.0){\rule[-0.200pt]{0.400pt}{1.204pt}}
\put(351.0,274.0){\usebox{\plotpoint}}
\put(352.0,271.0){\rule[-0.200pt]{0.400pt}{0.723pt}}
\put(351.67,270){\rule{0.400pt}{0.723pt}}
\multiput(351.17,271.50)(1.000,-1.500){2}{\rule{0.400pt}{0.361pt}}
\put(352.0,271.0){\rule[-0.200pt]{0.400pt}{0.482pt}}
\put(353,269.67){\rule{0.241pt}{0.400pt}}
\multiput(353.00,270.17)(0.500,-1.000){2}{\rule{0.120pt}{0.400pt}}
\put(353.0,270.0){\usebox{\plotpoint}}
\put(354.0,270.0){\usebox{\plotpoint}}
\put(354,266.67){\rule{0.241pt}{0.400pt}}
\multiput(354.00,267.17)(0.500,-1.000){2}{\rule{0.120pt}{0.400pt}}
\put(354.0,268.0){\rule[-0.200pt]{0.400pt}{0.723pt}}
\put(354.67,272){\rule{0.400pt}{1.204pt}}
\multiput(354.17,274.50)(1.000,-2.500){2}{\rule{0.400pt}{0.602pt}}
\put(355.0,267.0){\rule[-0.200pt]{0.400pt}{2.409pt}}
\put(355.67,277){\rule{0.400pt}{0.482pt}}
\multiput(355.17,278.00)(1.000,-1.000){2}{\rule{0.400pt}{0.241pt}}
\put(356.0,272.0){\rule[-0.200pt]{0.400pt}{1.686pt}}
\put(356.67,276){\rule{0.400pt}{0.964pt}}
\multiput(356.17,276.00)(1.000,2.000){2}{\rule{0.400pt}{0.482pt}}
\put(357.0,276.0){\usebox{\plotpoint}}
\put(358.0,278.0){\rule[-0.200pt]{0.400pt}{0.482pt}}
\put(357.67,282){\rule{0.400pt}{0.482pt}}
\multiput(357.17,283.00)(1.000,-1.000){2}{\rule{0.400pt}{0.241pt}}
\put(358.0,278.0){\rule[-0.200pt]{0.400pt}{1.445pt}}
\put(359.0,281.0){\usebox{\plotpoint}}
\put(359.0,281.0){\usebox{\plotpoint}}
\put(360.0,281.0){\rule[-0.200pt]{0.400pt}{2.168pt}}
\put(359.67,282){\rule{0.400pt}{3.614pt}}
\multiput(359.17,282.00)(1.000,7.500){2}{\rule{0.400pt}{1.807pt}}
\put(360.0,282.0){\rule[-0.200pt]{0.400pt}{1.927pt}}
\put(361,297){\usebox{\plotpoint}}
\put(360.67,282){\rule{0.400pt}{3.614pt}}
\multiput(360.17,289.50)(1.000,-7.500){2}{\rule{0.400pt}{1.807pt}}
\put(362.0,282.0){\rule[-0.200pt]{0.400pt}{1.686pt}}
\put(362,280.67){\rule{0.241pt}{0.400pt}}
\multiput(362.00,281.17)(0.500,-1.000){2}{\rule{0.120pt}{0.400pt}}
\put(362.0,282.0){\rule[-0.200pt]{0.400pt}{1.686pt}}
\put(363,281){\usebox{\plotpoint}}
\put(362.67,281){\rule{0.400pt}{0.723pt}}
\multiput(362.17,281.00)(1.000,1.500){2}{\rule{0.400pt}{0.361pt}}
\put(363.67,276){\rule{0.400pt}{0.482pt}}
\multiput(363.17,276.00)(1.000,1.000){2}{\rule{0.400pt}{0.241pt}}
\put(364.0,276.0){\rule[-0.200pt]{0.400pt}{1.927pt}}
\put(365.0,274.0){\rule[-0.200pt]{0.400pt}{0.964pt}}
\put(364.67,275){\rule{0.400pt}{0.723pt}}
\multiput(364.17,275.00)(1.000,1.500){2}{\rule{0.400pt}{0.361pt}}
\put(365.0,274.0){\usebox{\plotpoint}}
\put(366,278){\usebox{\plotpoint}}
\put(365.67,267){\rule{0.400pt}{2.650pt}}
\multiput(365.17,272.50)(1.000,-5.500){2}{\rule{0.400pt}{1.325pt}}
\put(367.0,267.0){\rule[-0.200pt]{0.400pt}{2.168pt}}
\put(366.67,255){\rule{0.400pt}{0.723pt}}
\multiput(366.17,256.50)(1.000,-1.500){2}{\rule{0.400pt}{0.361pt}}
\put(367.0,258.0){\rule[-0.200pt]{0.400pt}{4.336pt}}
\put(367.67,272){\rule{0.400pt}{4.095pt}}
\multiput(367.17,280.50)(1.000,-8.500){2}{\rule{0.400pt}{2.048pt}}
\put(368.0,255.0){\rule[-0.200pt]{0.400pt}{8.191pt}}
\put(368.67,289){\rule{0.400pt}{0.964pt}}
\multiput(368.17,289.00)(1.000,2.000){2}{\rule{0.400pt}{0.482pt}}
\put(369.0,272.0){\rule[-0.200pt]{0.400pt}{4.095pt}}
\put(369.67,283){\rule{0.400pt}{5.782pt}}
\multiput(369.17,283.00)(1.000,12.000){2}{\rule{0.400pt}{2.891pt}}
\put(370.0,283.0){\rule[-0.200pt]{0.400pt}{2.409pt}}
\put(370.67,297){\rule{0.400pt}{1.204pt}}
\multiput(370.17,297.00)(1.000,2.500){2}{\rule{0.400pt}{0.602pt}}
\put(371.0,297.0){\rule[-0.200pt]{0.400pt}{2.409pt}}
\put(372,290.67){\rule{0.241pt}{0.400pt}}
\multiput(372.00,290.17)(0.500,1.000){2}{\rule{0.120pt}{0.400pt}}
\put(372.0,291.0){\rule[-0.200pt]{0.400pt}{2.650pt}}
\put(373.0,292.0){\rule[-0.200pt]{0.400pt}{1.927pt}}
\put(372.67,296){\rule{0.400pt}{0.723pt}}
\multiput(372.17,296.00)(1.000,1.500){2}{\rule{0.400pt}{0.361pt}}
\put(373.0,296.0){\rule[-0.200pt]{0.400pt}{0.964pt}}
\put(373.67,297){\rule{0.400pt}{5.059pt}}
\multiput(373.17,297.00)(1.000,10.500){2}{\rule{0.400pt}{2.529pt}}
\put(374.0,297.0){\rule[-0.200pt]{0.400pt}{0.482pt}}
\put(375.0,304.0){\rule[-0.200pt]{0.400pt}{3.373pt}}
\put(375,321.67){\rule{0.241pt}{0.400pt}}
\multiput(375.00,321.17)(0.500,1.000){2}{\rule{0.120pt}{0.400pt}}
\put(375.0,304.0){\rule[-0.200pt]{0.400pt}{4.336pt}}
\put(375.67,311){\rule{0.400pt}{0.723pt}}
\multiput(375.17,311.00)(1.000,1.500){2}{\rule{0.400pt}{0.361pt}}
\put(376.0,311.0){\rule[-0.200pt]{0.400pt}{2.891pt}}
\put(377,314){\usebox{\plotpoint}}
\put(376.67,308){\rule{0.400pt}{7.709pt}}
\multiput(376.17,308.00)(1.000,16.000){2}{\rule{0.400pt}{3.854pt}}
\put(377.0,308.0){\rule[-0.200pt]{0.400pt}{1.445pt}}
\put(377.67,327){\rule{0.400pt}{0.482pt}}
\multiput(377.17,327.00)(1.000,1.000){2}{\rule{0.400pt}{0.241pt}}
\put(378.0,327.0){\rule[-0.200pt]{0.400pt}{3.132pt}}
\put(378.67,331){\rule{0.400pt}{6.745pt}}
\multiput(378.17,345.00)(1.000,-14.000){2}{\rule{0.400pt}{3.373pt}}
\put(379.0,329.0){\rule[-0.200pt]{0.400pt}{7.227pt}}
\put(380.0,331.0){\rule[-0.200pt]{0.400pt}{13.731pt}}
\put(380.0,388.0){\usebox{\plotpoint}}
\put(381.0,332.0){\rule[-0.200pt]{0.400pt}{13.490pt}}
\put(380.67,331){\rule{0.400pt}{6.986pt}}
\multiput(380.17,345.50)(1.000,-14.500){2}{\rule{0.400pt}{3.493pt}}
\put(381.0,332.0){\rule[-0.200pt]{0.400pt}{6.745pt}}
\put(382.0,330.0){\usebox{\plotpoint}}
\put(382.0,330.0){\usebox{\plotpoint}}
\put(383.0,330.0){\rule[-0.200pt]{0.400pt}{2.650pt}}
\put(382.67,312){\rule{0.400pt}{1.686pt}}
\multiput(382.17,312.00)(1.000,3.500){2}{\rule{0.400pt}{0.843pt}}
\put(383.0,312.0){\rule[-0.200pt]{0.400pt}{6.986pt}}
\put(384.0,313.0){\rule[-0.200pt]{0.400pt}{1.445pt}}
\put(384.0,313.0){\usebox{\plotpoint}}
\put(384.67,321){\rule{0.400pt}{0.482pt}}
\multiput(384.17,322.00)(1.000,-1.000){2}{\rule{0.400pt}{0.241pt}}
\put(385.0,313.0){\rule[-0.200pt]{0.400pt}{2.409pt}}
\put(386.0,302.0){\rule[-0.200pt]{0.400pt}{4.577pt}}
\put(385.67,290){\rule{0.400pt}{6.745pt}}
\multiput(385.17,304.00)(1.000,-14.000){2}{\rule{0.400pt}{3.373pt}}
\put(386.0,302.0){\rule[-0.200pt]{0.400pt}{3.854pt}}
\put(386.67,292){\rule{0.400pt}{2.409pt}}
\multiput(386.17,292.00)(1.000,5.000){2}{\rule{0.400pt}{1.204pt}}
\put(387.0,290.0){\rule[-0.200pt]{0.400pt}{0.482pt}}
\put(388.0,299.0){\rule[-0.200pt]{0.400pt}{0.723pt}}
\put(388.0,299.0){\usebox{\plotpoint}}
\put(388.67,300){\rule{0.400pt}{1.686pt}}
\multiput(388.17,303.50)(1.000,-3.500){2}{\rule{0.400pt}{0.843pt}}
\put(389.0,299.0){\rule[-0.200pt]{0.400pt}{1.927pt}}
\put(389.67,298){\rule{0.400pt}{3.614pt}}
\multiput(389.17,305.50)(1.000,-7.500){2}{\rule{0.400pt}{1.807pt}}
\put(390.0,300.0){\rule[-0.200pt]{0.400pt}{3.132pt}}
\put(390.67,300){\rule{0.400pt}{0.964pt}}
\multiput(390.17,302.00)(1.000,-2.000){2}{\rule{0.400pt}{0.482pt}}
\put(391.0,298.0){\rule[-0.200pt]{0.400pt}{1.445pt}}
\put(391.67,295){\rule{0.400pt}{1.445pt}}
\multiput(391.17,295.00)(1.000,3.000){2}{\rule{0.400pt}{0.723pt}}
\put(392.0,295.0){\rule[-0.200pt]{0.400pt}{1.204pt}}
\put(392.67,290){\rule{0.400pt}{0.723pt}}
\multiput(392.17,290.00)(1.000,1.500){2}{\rule{0.400pt}{0.361pt}}
\put(393.0,290.0){\rule[-0.200pt]{0.400pt}{2.650pt}}
\put(394.0,284.0){\rule[-0.200pt]{0.400pt}{2.168pt}}
\put(393.67,285){\rule{0.400pt}{0.964pt}}
\multiput(393.17,287.00)(1.000,-2.000){2}{\rule{0.400pt}{0.482pt}}
\put(394.0,284.0){\rule[-0.200pt]{0.400pt}{1.204pt}}
\put(395,285){\usebox{\plotpoint}}
\put(394.67,280){\rule{0.400pt}{1.204pt}}
\multiput(394.17,282.50)(1.000,-2.500){2}{\rule{0.400pt}{0.602pt}}
\put(396.0,280.0){\rule[-0.200pt]{0.400pt}{1.686pt}}
\put(396,272.67){\rule{0.241pt}{0.400pt}}
\multiput(396.00,272.17)(0.500,1.000){2}{\rule{0.120pt}{0.400pt}}
\put(396.0,273.0){\rule[-0.200pt]{0.400pt}{3.373pt}}
\put(396.67,278){\rule{0.400pt}{1.445pt}}
\multiput(396.17,281.00)(1.000,-3.000){2}{\rule{0.400pt}{0.723pt}}
\put(397.0,274.0){\rule[-0.200pt]{0.400pt}{2.409pt}}
\put(398.0,278.0){\rule[-0.200pt]{0.400pt}{1.927pt}}
\put(398,283.67){\rule{0.241pt}{0.400pt}}
\multiput(398.00,284.17)(0.500,-1.000){2}{\rule{0.120pt}{0.400pt}}
\put(398.0,285.0){\usebox{\plotpoint}}
\put(398.67,283){\rule{0.400pt}{1.686pt}}
\multiput(398.17,283.00)(1.000,3.500){2}{\rule{0.400pt}{0.843pt}}
\put(399.0,283.0){\usebox{\plotpoint}}
\put(400.0,290.0){\usebox{\plotpoint}}
\put(399.67,270){\rule{0.400pt}{2.891pt}}
\multiput(399.17,270.00)(1.000,6.000){2}{\rule{0.400pt}{1.445pt}}
\put(400.0,270.0){\rule[-0.200pt]{0.400pt}{5.059pt}}
\put(401,265.67){\rule{0.241pt}{0.400pt}}
\multiput(401.00,266.17)(0.500,-1.000){2}{\rule{0.120pt}{0.400pt}}
\put(401.0,267.0){\rule[-0.200pt]{0.400pt}{3.613pt}}
\put(402.0,266.0){\rule[-0.200pt]{0.400pt}{1.927pt}}
\put(401.67,270){\rule{0.400pt}{0.723pt}}
\multiput(401.17,271.50)(1.000,-1.500){2}{\rule{0.400pt}{0.361pt}}
\put(402.0,273.0){\usebox{\plotpoint}}
\put(402.67,254){\rule{0.400pt}{4.818pt}}
\multiput(402.17,264.00)(1.000,-10.000){2}{\rule{0.400pt}{2.409pt}}
\put(403.0,270.0){\rule[-0.200pt]{0.400pt}{0.964pt}}
\put(404,260.67){\rule{0.241pt}{0.400pt}}
\multiput(404.00,261.17)(0.500,-1.000){2}{\rule{0.120pt}{0.400pt}}
\put(404.0,254.0){\rule[-0.200pt]{0.400pt}{1.927pt}}
\put(404.67,243){\rule{0.400pt}{4.095pt}}
\multiput(404.17,243.00)(1.000,8.500){2}{\rule{0.400pt}{2.048pt}}
\put(405.0,243.0){\rule[-0.200pt]{0.400pt}{4.336pt}}
\put(405.67,227){\rule{0.400pt}{0.482pt}}
\multiput(405.17,227.00)(1.000,1.000){2}{\rule{0.400pt}{0.241pt}}
\put(406.0,227.0){\rule[-0.200pt]{0.400pt}{7.950pt}}
\put(407.0,229.0){\rule[-0.200pt]{0.400pt}{4.095pt}}
\put(406.67,238){\rule{0.400pt}{2.168pt}}
\multiput(406.17,238.00)(1.000,4.500){2}{\rule{0.400pt}{1.084pt}}
\put(407.0,238.0){\rule[-0.200pt]{0.400pt}{1.927pt}}
\put(408,247){\usebox{\plotpoint}}
\put(408,246.67){\rule{0.241pt}{0.400pt}}
\multiput(408.00,246.17)(0.500,1.000){2}{\rule{0.120pt}{0.400pt}}
\put(409.0,244.0){\rule[-0.200pt]{0.400pt}{0.964pt}}
\put(409,250.67){\rule{0.241pt}{0.400pt}}
\multiput(409.00,251.17)(0.500,-1.000){2}{\rule{0.120pt}{0.400pt}}
\put(409.0,244.0){\rule[-0.200pt]{0.400pt}{1.927pt}}
\put(410,249.67){\rule{0.241pt}{0.400pt}}
\multiput(410.00,249.17)(0.500,1.000){2}{\rule{0.120pt}{0.400pt}}
\put(410.0,250.0){\usebox{\plotpoint}}
\put(410.67,249){\rule{0.400pt}{0.482pt}}
\multiput(410.17,249.00)(1.000,1.000){2}{\rule{0.400pt}{0.241pt}}
\put(411.0,249.0){\rule[-0.200pt]{0.400pt}{0.482pt}}
\put(411.67,249){\rule{0.400pt}{0.482pt}}
\multiput(411.17,249.00)(1.000,1.000){2}{\rule{0.400pt}{0.241pt}}
\put(412.0,249.0){\rule[-0.200pt]{0.400pt}{0.482pt}}
\put(413,251){\usebox{\plotpoint}}
\put(413,251.67){\rule{0.241pt}{0.400pt}}
\multiput(413.00,252.17)(0.500,-1.000){2}{\rule{0.120pt}{0.400pt}}
\put(413.0,251.0){\rule[-0.200pt]{0.400pt}{0.482pt}}
\put(414.0,252.0){\rule[-0.200pt]{0.400pt}{0.482pt}}
\put(414.0,254.0){\usebox{\plotpoint}}
\put(415,252.67){\rule{0.241pt}{0.400pt}}
\multiput(415.00,252.17)(0.500,1.000){2}{\rule{0.120pt}{0.400pt}}
\put(415.0,253.0){\usebox{\plotpoint}}
\put(415.67,253){\rule{0.400pt}{0.723pt}}
\multiput(415.17,253.00)(1.000,1.500){2}{\rule{0.400pt}{0.361pt}}
\put(416.0,253.0){\usebox{\plotpoint}}
\put(417.0,255.0){\usebox{\plotpoint}}
\put(417.0,255.0){\usebox{\plotpoint}}
\put(417.67,255){\rule{0.400pt}{0.723pt}}
\multiput(417.17,256.50)(1.000,-1.500){2}{\rule{0.400pt}{0.361pt}}
\put(418.0,255.0){\rule[-0.200pt]{0.400pt}{0.723pt}}
\put(418.67,254){\rule{0.400pt}{1.927pt}}
\multiput(418.17,258.00)(1.000,-4.000){2}{\rule{0.400pt}{0.964pt}}
\put(419.0,255.0){\rule[-0.200pt]{0.400pt}{1.686pt}}
\put(419.67,255){\rule{0.400pt}{0.723pt}}
\multiput(419.17,256.50)(1.000,-1.500){2}{\rule{0.400pt}{0.361pt}}
\put(420.0,254.0){\rule[-0.200pt]{0.400pt}{0.964pt}}
\put(420.67,253){\rule{0.400pt}{0.723pt}}
\multiput(420.17,253.00)(1.000,1.500){2}{\rule{0.400pt}{0.361pt}}
\put(421.0,253.0){\rule[-0.200pt]{0.400pt}{0.482pt}}
\put(422.0,251.0){\rule[-0.200pt]{0.400pt}{1.204pt}}
\put(422.0,251.0){\usebox{\plotpoint}}
\put(423.0,251.0){\rule[-0.200pt]{0.400pt}{0.482pt}}
\put(422.67,252){\rule{0.400pt}{0.482pt}}
\multiput(422.17,252.00)(1.000,1.000){2}{\rule{0.400pt}{0.241pt}}
\put(423.0,252.0){\usebox{\plotpoint}}
\put(424,254){\usebox{\plotpoint}}
\put(423.67,252){\rule{0.400pt}{0.482pt}}
\multiput(423.17,253.00)(1.000,-1.000){2}{\rule{0.400pt}{0.241pt}}
\put(425.0,252.0){\usebox{\plotpoint}}
\put(425.0,253.0){\usebox{\plotpoint}}
\put(425.67,249){\rule{0.400pt}{0.723pt}}
\multiput(425.17,249.00)(1.000,1.500){2}{\rule{0.400pt}{0.361pt}}
\put(426.0,249.0){\rule[-0.200pt]{0.400pt}{0.964pt}}
\put(426.67,248){\rule{0.400pt}{0.482pt}}
\multiput(426.17,248.00)(1.000,1.000){2}{\rule{0.400pt}{0.241pt}}
\put(427.0,248.0){\rule[-0.200pt]{0.400pt}{0.964pt}}
\put(428,250){\usebox{\plotpoint}}
\put(428,248.67){\rule{0.241pt}{0.400pt}}
\multiput(428.00,249.17)(0.500,-1.000){2}{\rule{0.120pt}{0.400pt}}
\put(428.67,246){\rule{0.400pt}{0.964pt}}
\multiput(428.17,248.00)(1.000,-2.000){2}{\rule{0.400pt}{0.482pt}}
\put(429.0,249.0){\usebox{\plotpoint}}
\put(430.0,246.0){\usebox{\plotpoint}}
\put(430.0,247.0){\usebox{\plotpoint}}
\put(430.67,243){\rule{0.400pt}{0.964pt}}
\multiput(430.17,243.00)(1.000,2.000){2}{\rule{0.400pt}{0.482pt}}
\put(431.0,243.0){\rule[-0.200pt]{0.400pt}{0.964pt}}
\put(431.67,239){\rule{0.400pt}{1.204pt}}
\multiput(431.17,239.00)(1.000,2.500){2}{\rule{0.400pt}{0.602pt}}
\put(432.0,239.0){\rule[-0.200pt]{0.400pt}{1.927pt}}
\put(432.67,242){\rule{0.400pt}{0.482pt}}
\multiput(432.17,242.00)(1.000,1.000){2}{\rule{0.400pt}{0.241pt}}
\put(433.0,242.0){\rule[-0.200pt]{0.400pt}{0.482pt}}
\put(434,244){\usebox{\plotpoint}}
\put(434,244){\usebox{\plotpoint}}
\put(434.0,244.0){\usebox{\plotpoint}}
\put(435,244.67){\rule{0.241pt}{0.400pt}}
\multiput(435.00,245.17)(0.500,-1.000){2}{\rule{0.120pt}{0.400pt}}
\put(435.0,244.0){\rule[-0.200pt]{0.400pt}{0.482pt}}
\put(436,245){\usebox{\plotpoint}}
\put(436,245){\usebox{\plotpoint}}
\put(437,244.67){\rule{0.241pt}{0.400pt}}
\multiput(437.00,244.17)(0.500,1.000){2}{\rule{0.120pt}{0.400pt}}
\put(436.0,245.0){\usebox{\plotpoint}}
\put(438.0,245.0){\usebox{\plotpoint}}
\put(438.0,245.0){\usebox{\plotpoint}}
\put(438.0,246.0){\rule[-0.200pt]{0.964pt}{0.400pt}}
\put(442,245.67){\rule{0.241pt}{0.400pt}}
\multiput(442.00,246.17)(0.500,-1.000){2}{\rule{0.120pt}{0.400pt}}
\put(442.0,246.0){\usebox{\plotpoint}}
\put(443.0,246.0){\usebox{\plotpoint}}
\put(444,246.67){\rule{0.241pt}{0.400pt}}
\multiput(444.00,246.17)(0.500,1.000){2}{\rule{0.120pt}{0.400pt}}
\put(443.0,247.0){\usebox{\plotpoint}}
\put(445,248){\usebox{\plotpoint}}
\put(445,246.67){\rule{0.241pt}{0.400pt}}
\multiput(445.00,247.17)(0.500,-1.000){2}{\rule{0.120pt}{0.400pt}}
\put(446,247){\usebox{\plotpoint}}
\put(446,247){\usebox{\plotpoint}}
\put(446.0,247.0){\rule[-0.200pt]{0.482pt}{0.400pt}}
\put(448,245.67){\rule{0.241pt}{0.400pt}}
\multiput(448.00,245.17)(0.500,1.000){2}{\rule{0.120pt}{0.400pt}}
\put(448.0,246.0){\usebox{\plotpoint}}
\put(449,247){\usebox{\plotpoint}}
\put(450,245.67){\rule{0.241pt}{0.400pt}}
\multiput(450.00,246.17)(0.500,-1.000){2}{\rule{0.120pt}{0.400pt}}
\put(449.0,247.0){\usebox{\plotpoint}}
\put(451.0,246.0){\usebox{\plotpoint}}
\put(451.0,246.0){\usebox{\plotpoint}}
\put(457,244.67){\rule{0.241pt}{0.400pt}}
\multiput(457.00,245.17)(0.500,-1.000){2}{\rule{0.120pt}{0.400pt}}
\put(451.0,246.0){\rule[-0.200pt]{1.445pt}{0.400pt}}
\put(458,245){\usebox{\plotpoint}}
\put(458,244.67){\rule{0.241pt}{0.400pt}}
\multiput(458.00,244.17)(0.500,1.000){2}{\rule{0.120pt}{0.400pt}}
\put(459,246){\usebox{\plotpoint}}
\put(459,246){\usebox{\plotpoint}}
\put(459.0,246.0){\rule[-0.200pt]{10.600pt}{0.400pt}}
\end{picture}

%% file: fig2b1.latex
%%%  IMPORTANT REMARK
%%% 
%%%  This figure is in LaTeX format.
%%%  One must \input it in a LaTeX file and compile it.
%
%
%
% GNUPLOT: LaTeX picture
\setlength{\unitlength}{0.240900pt}
\ifx\plotpoint\undefined\newsavebox{\plotpoint}\fi
\begin{picture}(524,450)(0,0)
\font\gnuplot=cmr10 at 10pt
\gnuplot
\footnotesize
\sbox{\plotpoint}{\rule[-0.200pt]{0.400pt}{0.400pt}}%
\put(160.0,82.0){\rule[-0.200pt]{4.818pt}{0.400pt}}
\put(140,82){\makebox(0,0)[r]{-0.08}}
\put(483.0,82.0){\rule[-0.200pt]{4.818pt}{0.400pt}}
\put(160.0,164.0){\rule[-0.200pt]{4.818pt}{0.400pt}}
\put(140,164){\makebox(0,0)[r]{-0.04}}
\put(483.0,164.0){\rule[-0.200pt]{4.818pt}{0.400pt}}
\put(160.0,246.0){\rule[-0.200pt]{4.818pt}{0.400pt}}
\put(140,246){\makebox(0,0)[r]{0}}
\put(483.0,246.0){\rule[-0.200pt]{4.818pt}{0.400pt}}
\put(160.0,328.0){\rule[-0.200pt]{4.818pt}{0.400pt}}
\put(140,328){\makebox(0,0)[r]{0.04}}
\put(483.0,328.0){\rule[-0.200pt]{4.818pt}{0.400pt}}
\put(160.0,410.0){\rule[-0.200pt]{4.818pt}{0.400pt}}
\put(140,410){\makebox(0,0)[r]{0.08}}
\put(483.0,410.0){\rule[-0.200pt]{4.818pt}{0.400pt}}
\put(160.0,82.0){\rule[-0.200pt]{0.400pt}{4.818pt}}
\put(160,41){\makebox(0,0){25}}
\put(160.0,390.0){\rule[-0.200pt]{0.400pt}{4.818pt}}
\put(332.0,82.0){\rule[-0.200pt]{0.400pt}{4.818pt}}
\put(332,41){\makebox(0,0){75}}
\put(332.0,390.0){\rule[-0.200pt]{0.400pt}{4.818pt}}
\put(503.0,82.0){\rule[-0.200pt]{0.400pt}{4.818pt}}
\put(503,41){\makebox(0,0){125}}
\put(503.0,390.0){\rule[-0.200pt]{0.400pt}{4.818pt}}
\put(160.0,82.0){\rule[-0.200pt]{82.629pt}{0.400pt}}
\put(503.0,82.0){\rule[-0.200pt]{0.400pt}{79.015pt}}
\put(160.0,410.0){\rule[-0.200pt]{82.629pt}{0.400pt}}
\put(160.0,82.0){\rule[-0.200pt]{0.400pt}{79.015pt}}
\put(160,246){\usebox{\plotpoint}}
\put(215,245.67){\rule{0.723pt}{0.400pt}}
\multiput(215.00,245.17)(1.500,1.000){2}{\rule{0.361pt}{0.400pt}}
\put(218,245.67){\rule{0.964pt}{0.400pt}}
\multiput(218.00,246.17)(2.000,-1.000){2}{\rule{0.482pt}{0.400pt}}
\multiput(222.00,244.95)(0.462,-0.447){3}{\rule{0.500pt}{0.108pt}}
\multiput(222.00,245.17)(1.962,-3.000){2}{\rule{0.250pt}{0.400pt}}
\multiput(225.00,243.61)(0.685,0.447){3}{\rule{0.633pt}{0.108pt}}
\multiput(225.00,242.17)(2.685,3.000){2}{\rule{0.317pt}{0.400pt}}
\multiput(229.61,246.00)(0.447,0.909){3}{\rule{0.108pt}{0.767pt}}
\multiput(228.17,246.00)(3.000,3.409){2}{\rule{0.400pt}{0.383pt}}
\multiput(232.61,248.37)(0.447,-0.685){3}{\rule{0.108pt}{0.633pt}}
\multiput(231.17,249.69)(3.000,-2.685){2}{\rule{0.400pt}{0.317pt}}
\multiput(235.60,241.60)(0.468,-1.651){5}{\rule{0.113pt}{1.300pt}}
\multiput(234.17,244.30)(4.000,-9.302){2}{\rule{0.400pt}{0.650pt}}
\multiput(239.61,235.00)(0.447,3.141){3}{\rule{0.108pt}{2.100pt}}
\multiput(238.17,235.00)(3.000,10.641){2}{\rule{0.400pt}{1.050pt}}
\multiput(242.60,250.00)(0.468,1.212){5}{\rule{0.113pt}{1.000pt}}
\multiput(241.17,250.00)(4.000,6.924){2}{\rule{0.400pt}{0.500pt}}
\multiput(246.61,254.16)(0.447,-1.579){3}{\rule{0.108pt}{1.167pt}}
\multiput(245.17,256.58)(3.000,-5.579){2}{\rule{0.400pt}{0.583pt}}
\multiput(249.60,240.21)(0.468,-3.552){5}{\rule{0.113pt}{2.600pt}}
\multiput(248.17,245.60)(4.000,-19.604){2}{\rule{0.400pt}{1.300pt}}
\multiput(253.61,226.00)(0.447,10.062){3}{\rule{0.108pt}{6.233pt}}
\multiput(252.17,226.00)(3.000,33.062){2}{\rule{0.400pt}{3.117pt}}
\multiput(256.61,262.18)(0.447,-3.588){3}{\rule{0.108pt}{2.367pt}}
\multiput(255.17,267.09)(3.000,-12.088){2}{\rule{0.400pt}{1.183pt}}
\multiput(259.60,255.00)(0.468,2.528){5}{\rule{0.113pt}{1.900pt}}
\multiput(258.17,255.00)(4.000,14.056){2}{\rule{0.400pt}{0.950pt}}
\multiput(263.61,265.94)(0.447,-2.472){3}{\rule{0.108pt}{1.700pt}}
\multiput(262.17,269.47)(3.000,-8.472){2}{\rule{0.400pt}{0.850pt}}
\multiput(266.60,261.00)(0.468,9.986){5}{\rule{0.113pt}{7.000pt}}
\multiput(265.17,261.00)(4.000,54.471){2}{\rule{0.400pt}{3.500pt}}
\multiput(270.61,290.29)(0.447,-15.644){3}{\rule{0.108pt}{9.567pt}}
\multiput(269.17,310.14)(3.000,-51.144){2}{\rule{0.400pt}{4.783pt}}
\multiput(273.60,259.00)(0.468,8.962){5}{\rule{0.113pt}{6.300pt}}
\multiput(272.17,259.00)(4.000,48.924){2}{\rule{0.400pt}{3.150pt}}
\multiput(277.61,321.00)(0.447,4.927){3}{\rule{0.108pt}{3.167pt}}
\multiput(276.17,321.00)(3.000,16.427){2}{\rule{0.400pt}{1.583pt}}
\multiput(280.61,339.16)(0.447,-1.579){3}{\rule{0.108pt}{1.167pt}}
\multiput(279.17,341.58)(3.000,-5.579){2}{\rule{0.400pt}{0.583pt}}
\multiput(283.60,336.00)(0.468,5.014){5}{\rule{0.113pt}{3.600pt}}
\multiput(282.17,336.00)(4.000,27.528){2}{\rule{0.400pt}{1.800pt}}
\multiput(287.61,329.07)(0.447,-16.537){3}{\rule{0.108pt}{10.100pt}}
\multiput(286.17,350.04)(3.000,-54.037){2}{\rule{0.400pt}{5.050pt}}
\multiput(290.60,296.00)(0.468,1.505){5}{\rule{0.113pt}{1.200pt}}
\multiput(289.17,296.00)(4.000,8.509){2}{\rule{0.400pt}{0.600pt}}
\multiput(294.61,246.81)(0.447,-23.905){3}{\rule{0.108pt}{14.500pt}}
\multiput(293.17,276.90)(3.000,-77.905){2}{\rule{0.400pt}{7.250pt}}
\multiput(297.60,199.00)(0.468,2.090){5}{\rule{0.113pt}{1.600pt}}
\multiput(296.17,199.00)(4.000,11.679){2}{\rule{0.400pt}{0.800pt}}
\multiput(301.61,209.16)(0.447,-1.579){3}{\rule{0.108pt}{1.167pt}}
\multiput(300.17,211.58)(3.000,-5.579){2}{\rule{0.400pt}{0.583pt}}
\multiput(304.61,206.00)(0.447,16.983){3}{\rule{0.108pt}{10.367pt}}
\multiput(303.17,206.00)(3.000,55.483){2}{\rule{0.400pt}{5.183pt}}
\multiput(307.60,269.30)(0.468,-4.575){5}{\rule{0.113pt}{3.300pt}}
\multiput(306.17,276.15)(4.000,-25.151){2}{\rule{0.400pt}{1.650pt}}
\multiput(311.61,251.00)(0.447,2.695){3}{\rule{0.108pt}{1.833pt}}
\multiput(310.17,251.00)(3.000,9.195){2}{\rule{0.400pt}{0.917pt}}
\multiput(314.60,260.26)(0.468,-1.066){5}{\rule{0.113pt}{0.900pt}}
\multiput(313.17,262.13)(4.000,-6.132){2}{\rule{0.400pt}{0.450pt}}
\multiput(318.61,207.43)(0.447,-19.216){3}{\rule{0.108pt}{11.700pt}}
\multiput(317.17,231.72)(3.000,-62.716){2}{\rule{0.400pt}{5.850pt}}
\multiput(321.60,169.00)(0.468,2.090){5}{\rule{0.113pt}{1.600pt}}
\multiput(320.17,169.00)(4.000,11.679){2}{\rule{0.400pt}{0.800pt}}
\multiput(325.61,184.00)(0.447,8.053){3}{\rule{0.108pt}{5.033pt}}
\multiput(324.17,184.00)(3.000,26.553){2}{\rule{0.400pt}{2.517pt}}
\multiput(328.60,221.00)(0.468,7.207){5}{\rule{0.113pt}{5.100pt}}
\multiput(327.17,221.00)(4.000,39.415){2}{\rule{0.400pt}{2.550pt}}
\multiput(332.61,271.00)(0.447,8.053){3}{\rule{0.108pt}{5.033pt}}
\multiput(331.17,271.00)(3.000,26.553){2}{\rule{0.400pt}{2.517pt}}
\multiput(335.61,308.00)(0.447,3.141){3}{\rule{0.108pt}{2.100pt}}
\multiput(334.17,308.00)(3.000,10.641){2}{\rule{0.400pt}{1.050pt}}
\multiput(338.60,286.47)(0.468,-12.618){5}{\rule{0.113pt}{8.800pt}}
\multiput(337.17,304.74)(4.000,-68.735){2}{\rule{0.400pt}{4.400pt}}
\multiput(342.61,231.16)(0.447,-1.579){3}{\rule{0.108pt}{1.167pt}}
\multiput(341.17,233.58)(3.000,-5.579){2}{\rule{0.400pt}{0.583pt}}
\multiput(345.60,228.00)(0.468,1.797){5}{\rule{0.113pt}{1.400pt}}
\multiput(344.17,228.00)(4.000,10.094){2}{\rule{0.400pt}{0.700pt}}
\multiput(349.61,222.87)(0.447,-6.937){3}{\rule{0.108pt}{4.367pt}}
\multiput(348.17,231.94)(3.000,-22.937){2}{\rule{0.400pt}{2.183pt}}
\multiput(352.60,209.00)(0.468,11.155){5}{\rule{0.113pt}{7.800pt}}
\multiput(351.17,209.00)(4.000,60.811){2}{\rule{0.400pt}{3.900pt}}
\multiput(356.61,281.16)(0.447,-1.579){3}{\rule{0.108pt}{1.167pt}}
\multiput(355.17,283.58)(3.000,-5.579){2}{\rule{0.400pt}{0.583pt}}
\multiput(359.61,278.00)(0.447,3.141){3}{\rule{0.108pt}{2.100pt}}
\multiput(358.17,278.00)(3.000,10.641){2}{\rule{0.400pt}{1.050pt}}
\multiput(362.60,247.75)(0.468,-15.688){5}{\rule{0.113pt}{10.900pt}}
\multiput(361.17,270.38)(4.000,-85.377){2}{\rule{0.400pt}{5.450pt}}
\multiput(366.61,185.00)(0.447,2.248){3}{\rule{0.108pt}{1.567pt}}
\multiput(365.17,185.00)(3.000,7.748){2}{\rule{0.400pt}{0.783pt}}
\multiput(369.60,164.45)(0.468,-10.863){5}{\rule{0.113pt}{7.600pt}}
\multiput(368.17,180.23)(4.000,-59.226){2}{\rule{0.400pt}{3.800pt}}
\multiput(373.61,121.00)(0.447,7.607){3}{\rule{0.108pt}{4.767pt}}
\multiput(372.17,121.00)(3.000,25.107){2}{\rule{0.400pt}{2.383pt}}
\multiput(376.60,152.26)(0.468,-1.066){5}{\rule{0.113pt}{0.900pt}}
\multiput(375.17,154.13)(4.000,-6.132){2}{\rule{0.400pt}{0.450pt}}
\multiput(380.61,148.00)(0.447,4.927){3}{\rule{0.108pt}{3.167pt}}
\multiput(379.17,148.00)(3.000,16.427){2}{\rule{0.400pt}{1.583pt}}
\multiput(383.61,171.00)(0.447,13.635){3}{\rule{0.108pt}{8.367pt}}
\multiput(382.17,171.00)(3.000,44.635){2}{\rule{0.400pt}{4.183pt}}
\multiput(386.60,203.11)(0.468,-10.278){5}{\rule{0.113pt}{7.200pt}}
\multiput(385.17,218.06)(4.000,-56.056){2}{\rule{0.400pt}{3.600pt}}
\multiput(390.61,162.00)(0.447,15.197){3}{\rule{0.108pt}{9.300pt}}
\multiput(389.17,162.00)(3.000,49.697){2}{\rule{0.400pt}{4.650pt}}
\multiput(393.60,225.60)(0.468,-1.651){5}{\rule{0.113pt}{1.300pt}}
\multiput(392.17,228.30)(4.000,-9.302){2}{\rule{0.400pt}{0.650pt}}
\multiput(397.61,219.00)(0.447,3.811){3}{\rule{0.108pt}{2.500pt}}
\multiput(396.17,219.00)(3.000,12.811){2}{\rule{0.400pt}{1.250pt}}
\multiput(400.60,229.53)(0.468,-2.382){5}{\rule{0.113pt}{1.800pt}}
\multiput(399.17,233.26)(4.000,-13.264){2}{\rule{0.400pt}{0.900pt}}
\multiput(404.61,220.00)(0.447,10.062){3}{\rule{0.108pt}{6.233pt}}
\multiput(403.17,220.00)(3.000,33.062){2}{\rule{0.400pt}{3.117pt}}
\multiput(407.61,251.75)(0.447,-5.374){3}{\rule{0.108pt}{3.433pt}}
\multiput(406.17,258.87)(3.000,-17.874){2}{\rule{0.400pt}{1.717pt}}
\multiput(410.60,237.26)(0.468,-1.066){5}{\rule{0.113pt}{0.900pt}}
\multiput(409.17,239.13)(4.000,-6.132){2}{\rule{0.400pt}{0.450pt}}
\multiput(414.61,233.00)(0.447,1.802){3}{\rule{0.108pt}{1.300pt}}
\multiput(413.17,233.00)(3.000,6.302){2}{\rule{0.400pt}{0.650pt}}
\multiput(417.60,242.00)(0.468,2.090){5}{\rule{0.113pt}{1.600pt}}
\multiput(416.17,242.00)(4.000,11.679){2}{\rule{0.400pt}{0.800pt}}
\multiput(421.61,249.94)(0.447,-2.472){3}{\rule{0.108pt}{1.700pt}}
\multiput(420.17,253.47)(3.000,-8.472){2}{\rule{0.400pt}{0.850pt}}
\multiput(424.00,243.94)(0.481,-0.468){5}{\rule{0.500pt}{0.113pt}}
\multiput(424.00,244.17)(2.962,-4.000){2}{\rule{0.250pt}{0.400pt}}
\multiput(428.61,241.00)(0.447,0.909){3}{\rule{0.108pt}{0.767pt}}
\multiput(427.17,241.00)(3.000,3.409){2}{\rule{0.400pt}{0.383pt}}
\multiput(431.00,246.61)(0.462,0.447){3}{\rule{0.500pt}{0.108pt}}
\multiput(431.00,245.17)(1.962,3.000){2}{\rule{0.250pt}{0.400pt}}
\multiput(434.00,247.95)(0.685,-0.447){3}{\rule{0.633pt}{0.108pt}}
\multiput(434.00,248.17)(2.685,-3.000){2}{\rule{0.317pt}{0.400pt}}
\put(438,244.67){\rule{0.723pt}{0.400pt}}
\multiput(438.00,245.17)(1.500,-1.000){2}{\rule{0.361pt}{0.400pt}}
\put(441,244.67){\rule{0.964pt}{0.400pt}}
\multiput(441.00,244.17)(2.000,1.000){2}{\rule{0.482pt}{0.400pt}}
\put(160.0,246.0){\rule[-0.200pt]{13.249pt}{0.400pt}}
\put(445.0,246.0){\rule[-0.200pt]{13.972pt}{0.400pt}}
\end{picture}

%% file: fig2b2.latex
%%%  IMPORTANT REMARK
%%% 
%%%  This figure is in LaTeX format.
%%%  One must \input it in a LaTeX file and compile it.
%
%
%
% GNUPLOT: LaTeX picture
\setlength{\unitlength}{0.240900pt}
\ifx\plotpoint\undefined\newsavebox{\plotpoint}\fi
\begin{picture}(524,450)(0,0)
\font\gnuplot=cmr10 at 10pt
\gnuplot
\footnotesize
\sbox{\plotpoint}{\rule[-0.200pt]{0.400pt}{0.400pt}}%
\put(160.0,82.0){\rule[-0.200pt]{4.818pt}{0.400pt}}
\put(140,82){\makebox(0,0)[r]{-0.02}}
\put(483.0,82.0){\rule[-0.200pt]{4.818pt}{0.400pt}}
\put(160.0,164.0){\rule[-0.200pt]{4.818pt}{0.400pt}}
\put(140,164){\makebox(0,0)[r]{-0.01}}
\put(483.0,164.0){\rule[-0.200pt]{4.818pt}{0.400pt}}
\put(160.0,246.0){\rule[-0.200pt]{4.818pt}{0.400pt}}
\put(140,246){\makebox(0,0)[r]{0}}
\put(483.0,246.0){\rule[-0.200pt]{4.818pt}{0.400pt}}
\put(160.0,328.0){\rule[-0.200pt]{4.818pt}{0.400pt}}
\put(140,328){\makebox(0,0)[r]{0.01}}
\put(483.0,328.0){\rule[-0.200pt]{4.818pt}{0.400pt}}
\put(160.0,410.0){\rule[-0.200pt]{4.818pt}{0.400pt}}
\put(140,410){\makebox(0,0)[r]{0.02}}
\put(483.0,410.0){\rule[-0.200pt]{4.818pt}{0.400pt}}
\put(160.0,82.0){\rule[-0.200pt]{0.400pt}{4.818pt}}
\put(160,41){\makebox(0,0){200}}
\put(160.0,390.0){\rule[-0.200pt]{0.400pt}{4.818pt}}
\put(331.0,82.0){\rule[-0.200pt]{0.400pt}{4.818pt}}
\put(331,41){\makebox(0,0){350}}
\put(331.0,390.0){\rule[-0.200pt]{0.400pt}{4.818pt}}
\put(503.0,82.0){\rule[-0.200pt]{0.400pt}{4.818pt}}
\put(503,41){\makebox(0,0){500}}
\put(503.0,390.0){\rule[-0.200pt]{0.400pt}{4.818pt}}
\put(160.0,82.0){\rule[-0.200pt]{82.629pt}{0.400pt}}
\put(503.0,82.0){\rule[-0.200pt]{0.400pt}{79.015pt}}
\put(160.0,410.0){\rule[-0.200pt]{82.629pt}{0.400pt}}
\put(160.0,82.0){\rule[-0.200pt]{0.400pt}{79.015pt}}
\put(160,246){\usebox{\plotpoint}}
\put(165,245.67){\rule{0.241pt}{0.400pt}}
\multiput(165.00,245.17)(0.500,1.000){2}{\rule{0.120pt}{0.400pt}}
\put(166,245.67){\rule{0.241pt}{0.400pt}}
\multiput(166.00,246.17)(0.500,-1.000){2}{\rule{0.120pt}{0.400pt}}
\put(167,245.67){\rule{0.241pt}{0.400pt}}
\multiput(167.00,245.17)(0.500,1.000){2}{\rule{0.120pt}{0.400pt}}
\put(160.0,246.0){\rule[-0.200pt]{1.204pt}{0.400pt}}
\put(171,245.67){\rule{0.482pt}{0.400pt}}
\multiput(171.00,246.17)(1.000,-1.000){2}{\rule{0.241pt}{0.400pt}}
\put(168.0,247.0){\rule[-0.200pt]{0.723pt}{0.400pt}}
\put(173.67,244){\rule{0.400pt}{0.482pt}}
\multiput(173.17,245.00)(1.000,-1.000){2}{\rule{0.400pt}{0.241pt}}
\put(174.67,244){\rule{0.400pt}{0.482pt}}
\multiput(174.17,244.00)(1.000,1.000){2}{\rule{0.400pt}{0.241pt}}
\put(175.67,244){\rule{0.400pt}{0.482pt}}
\multiput(175.17,245.00)(1.000,-1.000){2}{\rule{0.400pt}{0.241pt}}
\put(173.0,246.0){\usebox{\plotpoint}}
\put(178,243.67){\rule{0.241pt}{0.400pt}}
\multiput(178.00,243.17)(0.500,1.000){2}{\rule{0.120pt}{0.400pt}}
\put(179,243.67){\rule{0.482pt}{0.400pt}}
\multiput(179.00,244.17)(1.000,-1.000){2}{\rule{0.241pt}{0.400pt}}
\put(180.67,244){\rule{0.400pt}{0.482pt}}
\multiput(180.17,244.00)(1.000,1.000){2}{\rule{0.400pt}{0.241pt}}
\put(182,244.67){\rule{0.241pt}{0.400pt}}
\multiput(182.00,245.17)(0.500,-1.000){2}{\rule{0.120pt}{0.400pt}}
\put(182.67,245){\rule{0.400pt}{0.964pt}}
\multiput(182.17,245.00)(1.000,2.000){2}{\rule{0.400pt}{0.482pt}}
\put(183.67,245){\rule{0.400pt}{0.964pt}}
\multiput(183.17,247.00)(1.000,-2.000){2}{\rule{0.400pt}{0.482pt}}
\put(184.67,245){\rule{0.400pt}{1.204pt}}
\multiput(184.17,245.00)(1.000,2.500){2}{\rule{0.400pt}{0.602pt}}
\put(186,249.67){\rule{0.241pt}{0.400pt}}
\multiput(186.00,249.17)(0.500,1.000){2}{\rule{0.120pt}{0.400pt}}
\put(187,249.17){\rule{0.482pt}{0.400pt}}
\multiput(187.00,250.17)(1.000,-2.000){2}{\rule{0.241pt}{0.400pt}}
\put(188.67,249){\rule{0.400pt}{0.723pt}}
\multiput(188.17,249.00)(1.000,1.500){2}{\rule{0.400pt}{0.361pt}}
\put(189.67,246){\rule{0.400pt}{1.445pt}}
\multiput(189.17,249.00)(1.000,-3.000){2}{\rule{0.400pt}{0.723pt}}
\put(191,245.67){\rule{0.241pt}{0.400pt}}
\multiput(191.00,245.17)(0.500,1.000){2}{\rule{0.120pt}{0.400pt}}
\put(191.67,239){\rule{0.400pt}{1.927pt}}
\multiput(191.17,243.00)(1.000,-4.000){2}{\rule{0.400pt}{0.964pt}}
\put(192.67,239){\rule{0.400pt}{1.445pt}}
\multiput(192.17,239.00)(1.000,3.000){2}{\rule{0.400pt}{0.723pt}}
\put(193.67,239){\rule{0.400pt}{1.445pt}}
\multiput(193.17,242.00)(1.000,-3.000){2}{\rule{0.400pt}{0.723pt}}
\put(177.0,244.0){\usebox{\plotpoint}}
\put(196.67,239){\rule{0.400pt}{0.964pt}}
\multiput(196.17,239.00)(1.000,2.000){2}{\rule{0.400pt}{0.482pt}}
\put(197.67,239){\rule{0.400pt}{0.964pt}}
\multiput(197.17,241.00)(1.000,-2.000){2}{\rule{0.400pt}{0.482pt}}
\put(198.67,239){\rule{0.400pt}{1.686pt}}
\multiput(198.17,239.00)(1.000,3.500){2}{\rule{0.400pt}{0.843pt}}
\put(199.67,241){\rule{0.400pt}{1.204pt}}
\multiput(199.17,243.50)(1.000,-2.500){2}{\rule{0.400pt}{0.602pt}}
\put(200.67,241){\rule{0.400pt}{3.614pt}}
\multiput(200.17,241.00)(1.000,7.500){2}{\rule{0.400pt}{1.807pt}}
\put(201.67,247){\rule{0.400pt}{2.168pt}}
\multiput(201.17,251.50)(1.000,-4.500){2}{\rule{0.400pt}{1.084pt}}
\put(203.17,247){\rule{0.400pt}{2.300pt}}
\multiput(202.17,247.00)(2.000,6.226){2}{\rule{0.400pt}{1.150pt}}
\put(195.0,239.0){\rule[-0.200pt]{0.482pt}{0.400pt}}
\put(206,256.67){\rule{0.241pt}{0.400pt}}
\multiput(206.00,257.17)(0.500,-1.000){2}{\rule{0.120pt}{0.400pt}}
\put(206.67,257){\rule{0.400pt}{1.686pt}}
\multiput(206.17,257.00)(1.000,3.500){2}{\rule{0.400pt}{0.843pt}}
\put(207.67,248){\rule{0.400pt}{3.854pt}}
\multiput(207.17,256.00)(1.000,-8.000){2}{\rule{0.400pt}{1.927pt}}
\put(209,247.67){\rule{0.241pt}{0.400pt}}
\multiput(209.00,247.17)(0.500,1.000){2}{\rule{0.120pt}{0.400pt}}
\put(209.67,236){\rule{0.400pt}{3.132pt}}
\multiput(209.17,242.50)(1.000,-6.500){2}{\rule{0.400pt}{1.566pt}}
\put(211.17,236){\rule{0.400pt}{0.900pt}}
\multiput(210.17,236.00)(2.000,2.132){2}{\rule{0.400pt}{0.450pt}}
\put(212.67,237){\rule{0.400pt}{0.723pt}}
\multiput(212.17,238.50)(1.000,-1.500){2}{\rule{0.400pt}{0.361pt}}
\put(213.67,237){\rule{0.400pt}{0.723pt}}
\multiput(213.17,237.00)(1.000,1.500){2}{\rule{0.400pt}{0.361pt}}
\put(214.67,240){\rule{0.400pt}{3.132pt}}
\multiput(214.17,240.00)(1.000,6.500){2}{\rule{0.400pt}{1.566pt}}
\put(215.67,247){\rule{0.400pt}{1.445pt}}
\multiput(215.17,250.00)(1.000,-3.000){2}{\rule{0.400pt}{0.723pt}}
\put(216.67,247){\rule{0.400pt}{2.168pt}}
\multiput(216.17,247.00)(1.000,4.500){2}{\rule{0.400pt}{1.084pt}}
\put(217.67,237){\rule{0.400pt}{4.577pt}}
\multiput(217.17,246.50)(1.000,-9.500){2}{\rule{0.400pt}{2.289pt}}
\put(219.17,237){\rule{0.400pt}{4.900pt}}
\multiput(218.17,237.00)(2.000,13.830){2}{\rule{0.400pt}{2.450pt}}
\put(221,259.67){\rule{0.241pt}{0.400pt}}
\multiput(221.00,260.17)(0.500,-1.000){2}{\rule{0.120pt}{0.400pt}}
\put(221.67,260){\rule{0.400pt}{1.927pt}}
\multiput(221.17,260.00)(1.000,4.000){2}{\rule{0.400pt}{0.964pt}}
\put(222.67,263){\rule{0.400pt}{1.204pt}}
\multiput(222.17,265.50)(1.000,-2.500){2}{\rule{0.400pt}{0.602pt}}
\put(223.67,263){\rule{0.400pt}{3.132pt}}
\multiput(223.17,263.00)(1.000,6.500){2}{\rule{0.400pt}{1.566pt}}
\put(205.0,258.0){\usebox{\plotpoint}}
\put(225.67,255){\rule{0.400pt}{5.059pt}}
\multiput(225.17,265.50)(1.000,-10.500){2}{\rule{0.400pt}{2.529pt}}
\put(227.17,245){\rule{0.400pt}{2.100pt}}
\multiput(226.17,250.64)(2.000,-5.641){2}{\rule{0.400pt}{1.050pt}}
\put(228.67,245){\rule{0.400pt}{5.300pt}}
\multiput(228.17,245.00)(1.000,11.000){2}{\rule{0.400pt}{2.650pt}}
\put(229.67,254){\rule{0.400pt}{3.132pt}}
\multiput(229.17,260.50)(1.000,-6.500){2}{\rule{0.400pt}{1.566pt}}
\put(230.67,254){\rule{0.400pt}{5.300pt}}
\multiput(230.17,254.00)(1.000,11.000){2}{\rule{0.400pt}{2.650pt}}
\put(231.67,274){\rule{0.400pt}{0.482pt}}
\multiput(231.17,275.00)(1.000,-1.000){2}{\rule{0.400pt}{0.241pt}}
\put(232.67,274){\rule{0.400pt}{8.913pt}}
\multiput(232.17,274.00)(1.000,18.500){2}{\rule{0.400pt}{4.457pt}}
\put(233.67,297){\rule{0.400pt}{3.373pt}}
\multiput(233.17,304.00)(1.000,-7.000){2}{\rule{0.400pt}{1.686pt}}
\put(235,295.67){\rule{0.482pt}{0.400pt}}
\multiput(235.00,296.17)(1.000,-1.000){2}{\rule{0.241pt}{0.400pt}}
\put(236.67,273){\rule{0.400pt}{5.541pt}}
\multiput(236.17,284.50)(1.000,-11.500){2}{\rule{0.400pt}{2.770pt}}
\put(237.67,273){\rule{0.400pt}{3.854pt}}
\multiput(237.17,273.00)(1.000,8.000){2}{\rule{0.400pt}{1.927pt}}
\put(238.67,289){\rule{0.400pt}{3.373pt}}
\multiput(238.17,289.00)(1.000,7.000){2}{\rule{0.400pt}{1.686pt}}
\put(239.67,291){\rule{0.400pt}{2.891pt}}
\multiput(239.17,297.00)(1.000,-6.000){2}{\rule{0.400pt}{1.445pt}}
\put(241,290.67){\rule{0.241pt}{0.400pt}}
\multiput(241.00,290.17)(0.500,1.000){2}{\rule{0.120pt}{0.400pt}}
\put(241.67,292){\rule{0.400pt}{4.818pt}}
\multiput(241.17,292.00)(1.000,10.000){2}{\rule{0.400pt}{2.409pt}}
\put(243.17,287){\rule{0.400pt}{5.100pt}}
\multiput(242.17,301.41)(2.000,-14.415){2}{\rule{0.400pt}{2.550pt}}
\put(244.67,287){\rule{0.400pt}{0.964pt}}
\multiput(244.17,287.00)(1.000,2.000){2}{\rule{0.400pt}{0.482pt}}
\put(245.67,276){\rule{0.400pt}{3.614pt}}
\multiput(245.17,283.50)(1.000,-7.500){2}{\rule{0.400pt}{1.807pt}}
\put(246.67,276){\rule{0.400pt}{13.250pt}}
\multiput(246.17,276.00)(1.000,27.500){2}{\rule{0.400pt}{6.625pt}}
\put(247.67,327){\rule{0.400pt}{0.964pt}}
\multiput(247.17,329.00)(1.000,-2.000){2}{\rule{0.400pt}{0.482pt}}
\put(248.67,327){\rule{0.400pt}{4.818pt}}
\multiput(248.17,327.00)(1.000,10.000){2}{\rule{0.400pt}{2.409pt}}
\put(249.67,319){\rule{0.400pt}{6.745pt}}
\multiput(249.17,333.00)(1.000,-14.000){2}{\rule{0.400pt}{3.373pt}}
\put(251.17,319){\rule{0.400pt}{8.300pt}}
\multiput(250.17,319.00)(2.000,23.773){2}{\rule{0.400pt}{4.150pt}}
\put(252.67,327){\rule{0.400pt}{7.950pt}}
\multiput(252.17,343.50)(1.000,-16.500){2}{\rule{0.400pt}{3.975pt}}
\put(253.67,327){\rule{0.400pt}{1.445pt}}
\multiput(253.17,327.00)(1.000,3.000){2}{\rule{0.400pt}{0.723pt}}
\put(254.67,333){\rule{0.400pt}{6.023pt}}
\multiput(254.17,333.00)(1.000,12.500){2}{\rule{0.400pt}{3.011pt}}
\put(255.67,358){\rule{0.400pt}{2.409pt}}
\multiput(255.17,358.00)(1.000,5.000){2}{\rule{0.400pt}{1.204pt}}
\put(256.67,350){\rule{0.400pt}{4.336pt}}
\multiput(256.17,359.00)(1.000,-9.000){2}{\rule{0.400pt}{2.168pt}}
\put(257.67,346){\rule{0.400pt}{0.964pt}}
\multiput(257.17,348.00)(1.000,-2.000){2}{\rule{0.400pt}{0.482pt}}
\put(259.17,346){\rule{0.400pt}{3.100pt}}
\multiput(258.17,346.00)(2.000,8.566){2}{\rule{0.400pt}{1.550pt}}
\put(260.67,323){\rule{0.400pt}{9.154pt}}
\multiput(260.17,342.00)(1.000,-19.000){2}{\rule{0.400pt}{4.577pt}}
\put(225.0,276.0){\usebox{\plotpoint}}
\put(262.67,323){\rule{0.400pt}{0.723pt}}
\multiput(262.17,323.00)(1.000,1.500){2}{\rule{0.400pt}{0.361pt}}
\put(263.67,326){\rule{0.400pt}{12.768pt}}
\multiput(263.17,326.00)(1.000,26.500){2}{\rule{0.400pt}{6.384pt}}
\put(264.67,298){\rule{0.400pt}{19.513pt}}
\multiput(264.17,338.50)(1.000,-40.500){2}{\rule{0.400pt}{9.756pt}}
\put(265.67,298){\rule{0.400pt}{3.132pt}}
\multiput(265.17,298.00)(1.000,6.500){2}{\rule{0.400pt}{1.566pt}}
\put(267.17,286){\rule{0.400pt}{5.100pt}}
\multiput(266.17,300.41)(2.000,-14.415){2}{\rule{0.400pt}{2.550pt}}
\put(268.67,286){\rule{0.400pt}{7.468pt}}
\multiput(268.17,286.00)(1.000,15.500){2}{\rule{0.400pt}{3.734pt}}
\put(269.67,210){\rule{0.400pt}{25.776pt}}
\multiput(269.17,263.50)(1.000,-53.500){2}{\rule{0.400pt}{12.888pt}}
\put(270.67,210){\rule{0.400pt}{13.731pt}}
\multiput(270.17,210.00)(1.000,28.500){2}{\rule{0.400pt}{6.866pt}}
\put(271.67,251){\rule{0.400pt}{3.854pt}}
\multiput(271.17,259.00)(1.000,-8.000){2}{\rule{0.400pt}{1.927pt}}
\put(272.67,251){\rule{0.400pt}{5.300pt}}
\multiput(272.17,251.00)(1.000,11.000){2}{\rule{0.400pt}{2.650pt}}
\put(273.67,190){\rule{0.400pt}{19.995pt}}
\multiput(273.17,231.50)(1.000,-41.500){2}{\rule{0.400pt}{9.997pt}}
\put(275.17,190){\rule{0.400pt}{8.100pt}}
\multiput(274.17,190.00)(2.000,23.188){2}{\rule{0.400pt}{4.050pt}}
\put(276.67,205){\rule{0.400pt}{6.023pt}}
\multiput(276.17,217.50)(1.000,-12.500){2}{\rule{0.400pt}{3.011pt}}
\put(277.67,186){\rule{0.400pt}{4.577pt}}
\multiput(277.17,195.50)(1.000,-9.500){2}{\rule{0.400pt}{2.289pt}}
\put(278.67,182){\rule{0.400pt}{0.964pt}}
\multiput(278.17,184.00)(1.000,-2.000){2}{\rule{0.400pt}{0.482pt}}
\put(279.67,182){\rule{0.400pt}{4.818pt}}
\multiput(279.17,182.00)(1.000,10.000){2}{\rule{0.400pt}{2.409pt}}
\put(280.67,202){\rule{0.400pt}{8.191pt}}
\multiput(280.17,202.00)(1.000,17.000){2}{\rule{0.400pt}{4.095pt}}
\put(281.67,205){\rule{0.400pt}{7.468pt}}
\multiput(281.17,220.50)(1.000,-15.500){2}{\rule{0.400pt}{3.734pt}}
\put(283.17,205){\rule{0.400pt}{10.900pt}}
\multiput(282.17,205.00)(2.000,31.377){2}{\rule{0.400pt}{5.450pt}}
\put(284.67,236){\rule{0.400pt}{5.541pt}}
\multiput(284.17,247.50)(1.000,-11.500){2}{\rule{0.400pt}{2.770pt}}
\put(285.67,236){\rule{0.400pt}{3.854pt}}
\multiput(285.17,236.00)(1.000,8.000){2}{\rule{0.400pt}{1.927pt}}
\put(286.67,229){\rule{0.400pt}{5.541pt}}
\multiput(286.17,240.50)(1.000,-11.500){2}{\rule{0.400pt}{2.770pt}}
\put(287.67,229){\rule{0.400pt}{10.359pt}}
\multiput(287.17,229.00)(1.000,21.500){2}{\rule{0.400pt}{5.179pt}}
\put(288.67,263){\rule{0.400pt}{2.168pt}}
\multiput(288.17,267.50)(1.000,-4.500){2}{\rule{0.400pt}{1.084pt}}
\put(290,262.67){\rule{0.241pt}{0.400pt}}
\multiput(290.00,262.17)(0.500,1.000){2}{\rule{0.120pt}{0.400pt}}
\put(291.17,264){\rule{0.400pt}{2.500pt}}
\multiput(290.17,264.00)(2.000,6.811){2}{\rule{0.400pt}{1.250pt}}
\put(292.67,276){\rule{0.400pt}{0.482pt}}
\multiput(292.17,276.00)(1.000,1.000){2}{\rule{0.400pt}{0.241pt}}
\put(293.67,244){\rule{0.400pt}{8.191pt}}
\multiput(293.17,261.00)(1.000,-17.000){2}{\rule{0.400pt}{4.095pt}}
\put(294.67,244){\rule{0.400pt}{0.482pt}}
\multiput(294.17,244.00)(1.000,1.000){2}{\rule{0.400pt}{0.241pt}}
\put(295.67,246){\rule{0.400pt}{8.913pt}}
\multiput(295.17,246.00)(1.000,18.500){2}{\rule{0.400pt}{4.457pt}}
\put(296.67,252){\rule{0.400pt}{7.468pt}}
\multiput(296.17,267.50)(1.000,-15.500){2}{\rule{0.400pt}{3.734pt}}
\put(297.67,252){\rule{0.400pt}{2.409pt}}
\multiput(297.17,252.00)(1.000,5.000){2}{\rule{0.400pt}{1.204pt}}
\put(299.17,255){\rule{0.400pt}{1.500pt}}
\multiput(298.17,258.89)(2.000,-3.887){2}{\rule{0.400pt}{0.750pt}}
\put(300.67,255){\rule{0.400pt}{3.614pt}}
\multiput(300.17,255.00)(1.000,7.500){2}{\rule{0.400pt}{1.807pt}}
\put(301.67,193){\rule{0.400pt}{18.549pt}}
\multiput(301.17,231.50)(1.000,-38.500){2}{\rule{0.400pt}{9.275pt}}
\put(302.67,193){\rule{0.400pt}{10.118pt}}
\multiput(302.17,193.00)(1.000,21.000){2}{\rule{0.400pt}{5.059pt}}
\put(303.67,197){\rule{0.400pt}{9.154pt}}
\multiput(303.17,216.00)(1.000,-19.000){2}{\rule{0.400pt}{4.577pt}}
\put(304.67,197){\rule{0.400pt}{4.818pt}}
\multiput(304.17,197.00)(1.000,10.000){2}{\rule{0.400pt}{2.409pt}}
\put(305.67,163){\rule{0.400pt}{13.009pt}}
\multiput(305.17,190.00)(1.000,-27.000){2}{\rule{0.400pt}{6.504pt}}
\put(307.17,163){\rule{0.400pt}{5.900pt}}
\multiput(306.17,163.00)(2.000,16.754){2}{\rule{0.400pt}{2.950pt}}
\put(308.67,184){\rule{0.400pt}{1.927pt}}
\multiput(308.17,188.00)(1.000,-4.000){2}{\rule{0.400pt}{0.964pt}}
\put(309.67,152){\rule{0.400pt}{7.709pt}}
\multiput(309.17,168.00)(1.000,-16.000){2}{\rule{0.400pt}{3.854pt}}
\put(310.67,152){\rule{0.400pt}{8.672pt}}
\multiput(310.17,152.00)(1.000,18.000){2}{\rule{0.400pt}{4.336pt}}
\put(311.67,188){\rule{0.400pt}{2.891pt}}
\multiput(311.17,188.00)(1.000,6.000){2}{\rule{0.400pt}{1.445pt}}
\put(312.67,200){\rule{0.400pt}{6.986pt}}
\multiput(312.17,200.00)(1.000,14.500){2}{\rule{0.400pt}{3.493pt}}
\put(313.67,172){\rule{0.400pt}{13.731pt}}
\multiput(313.17,200.50)(1.000,-28.500){2}{\rule{0.400pt}{6.866pt}}
\put(315.17,172){\rule{0.400pt}{18.700pt}}
\multiput(314.17,172.00)(2.000,54.187){2}{\rule{0.400pt}{9.350pt}}
\put(316.67,237){\rule{0.400pt}{6.745pt}}
\multiput(316.17,251.00)(1.000,-14.000){2}{\rule{0.400pt}{3.373pt}}
\put(317.67,237){\rule{0.400pt}{4.336pt}}
\multiput(317.17,237.00)(1.000,9.000){2}{\rule{0.400pt}{2.168pt}}
\put(318.67,227){\rule{0.400pt}{6.745pt}}
\multiput(318.17,241.00)(1.000,-14.000){2}{\rule{0.400pt}{3.373pt}}
\put(319.67,227){\rule{0.400pt}{22.404pt}}
\multiput(319.17,227.00)(1.000,46.500){2}{\rule{0.400pt}{11.202pt}}
\put(320.67,263){\rule{0.400pt}{13.731pt}}
\multiput(320.17,291.50)(1.000,-28.500){2}{\rule{0.400pt}{6.866pt}}
\put(321.67,263){\rule{0.400pt}{6.986pt}}
\multiput(321.17,263.00)(1.000,14.500){2}{\rule{0.400pt}{3.493pt}}
\put(323.17,292){\rule{0.400pt}{2.500pt}}
\multiput(322.17,292.00)(2.000,6.811){2}{\rule{0.400pt}{1.250pt}}
\put(324.67,304){\rule{0.400pt}{8.672pt}}
\multiput(324.17,304.00)(1.000,18.000){2}{\rule{0.400pt}{4.336pt}}
\put(325.67,308){\rule{0.400pt}{7.709pt}}
\multiput(325.17,324.00)(1.000,-16.000){2}{\rule{0.400pt}{3.854pt}}
\put(326.67,300){\rule{0.400pt}{1.927pt}}
\multiput(326.17,304.00)(1.000,-4.000){2}{\rule{0.400pt}{0.964pt}}
\put(327.67,300){\rule{0.400pt}{6.986pt}}
\multiput(327.17,300.00)(1.000,14.500){2}{\rule{0.400pt}{3.493pt}}
\put(328.67,275){\rule{0.400pt}{13.009pt}}
\multiput(328.17,302.00)(1.000,-27.000){2}{\rule{0.400pt}{6.504pt}}
\put(329.67,275){\rule{0.400pt}{4.818pt}}
\multiput(329.17,275.00)(1.000,10.000){2}{\rule{0.400pt}{2.409pt}}
\put(331.17,257){\rule{0.400pt}{7.700pt}}
\multiput(330.17,279.02)(2.000,-22.018){2}{\rule{0.400pt}{3.850pt}}
\put(332.67,257){\rule{0.400pt}{10.118pt}}
\multiput(332.17,257.00)(1.000,21.000){2}{\rule{0.400pt}{5.059pt}}
\put(333.67,222){\rule{0.400pt}{18.549pt}}
\multiput(333.17,260.50)(1.000,-38.500){2}{\rule{0.400pt}{9.275pt}}
\put(334.67,222){\rule{0.400pt}{3.614pt}}
\multiput(334.17,222.00)(1.000,7.500){2}{\rule{0.400pt}{1.807pt}}
\put(335.67,230){\rule{0.400pt}{1.686pt}}
\multiput(335.17,233.50)(1.000,-3.500){2}{\rule{0.400pt}{0.843pt}}
\put(336.67,230){\rule{0.400pt}{2.409pt}}
\multiput(336.17,230.00)(1.000,5.000){2}{\rule{0.400pt}{1.204pt}}
\put(338.17,209){\rule{0.400pt}{6.300pt}}
\multiput(337.17,226.92)(2.000,-17.924){2}{\rule{0.400pt}{3.150pt}}
\put(339.67,209){\rule{0.400pt}{8.913pt}}
\multiput(339.17,209.00)(1.000,18.500){2}{\rule{0.400pt}{4.457pt}}
\put(340.67,246){\rule{0.400pt}{0.482pt}}
\multiput(340.17,246.00)(1.000,1.000){2}{\rule{0.400pt}{0.241pt}}
\put(341.67,214){\rule{0.400pt}{8.191pt}}
\multiput(341.17,231.00)(1.000,-17.000){2}{\rule{0.400pt}{4.095pt}}
\put(342.67,214){\rule{0.400pt}{0.482pt}}
\multiput(342.17,214.00)(1.000,1.000){2}{\rule{0.400pt}{0.241pt}}
\put(343.67,216){\rule{0.400pt}{2.891pt}}
\multiput(343.17,216.00)(1.000,6.000){2}{\rule{0.400pt}{1.445pt}}
\put(345,227.67){\rule{0.241pt}{0.400pt}}
\multiput(345.00,227.17)(0.500,1.000){2}{\rule{0.120pt}{0.400pt}}
\put(346.17,220){\rule{0.400pt}{1.900pt}}
\multiput(345.17,225.06)(2.000,-5.056){2}{\rule{0.400pt}{0.950pt}}
\put(347.67,220){\rule{0.400pt}{10.359pt}}
\multiput(347.17,220.00)(1.000,21.500){2}{\rule{0.400pt}{5.179pt}}
\put(348.67,240){\rule{0.400pt}{5.541pt}}
\multiput(348.17,251.50)(1.000,-11.500){2}{\rule{0.400pt}{2.770pt}}
\put(349.67,240){\rule{0.400pt}{3.854pt}}
\multiput(349.17,240.00)(1.000,8.000){2}{\rule{0.400pt}{1.927pt}}
\put(350.67,233){\rule{0.400pt}{5.541pt}}
\multiput(350.17,244.50)(1.000,-11.500){2}{\rule{0.400pt}{2.770pt}}
\put(351.67,233){\rule{0.400pt}{13.009pt}}
\multiput(351.17,233.00)(1.000,27.000){2}{\rule{0.400pt}{6.504pt}}
\put(352.67,256){\rule{0.400pt}{7.468pt}}
\multiput(352.17,271.50)(1.000,-15.500){2}{\rule{0.400pt}{3.734pt}}
\put(354.17,256){\rule{0.400pt}{6.900pt}}
\multiput(353.17,256.00)(2.000,19.679){2}{\rule{0.400pt}{3.450pt}}
\put(355.67,290){\rule{0.400pt}{4.818pt}}
\multiput(355.17,290.00)(1.000,10.000){2}{\rule{0.400pt}{2.409pt}}
\put(356.67,306){\rule{0.400pt}{0.964pt}}
\multiput(356.17,308.00)(1.000,-2.000){2}{\rule{0.400pt}{0.482pt}}
\put(357.67,287){\rule{0.400pt}{4.577pt}}
\multiput(357.17,296.50)(1.000,-9.500){2}{\rule{0.400pt}{2.289pt}}
\put(358.67,262){\rule{0.400pt}{6.023pt}}
\multiput(358.17,274.50)(1.000,-12.500){2}{\rule{0.400pt}{3.011pt}}
\put(359.67,262){\rule{0.400pt}{9.636pt}}
\multiput(359.17,262.00)(1.000,20.000){2}{\rule{0.400pt}{4.818pt}}
\put(360.67,219){\rule{0.400pt}{19.995pt}}
\multiput(360.17,260.50)(1.000,-41.500){2}{\rule{0.400pt}{9.997pt}}
\put(362.17,219){\rule{0.400pt}{4.500pt}}
\multiput(361.17,219.00)(2.000,12.660){2}{\rule{0.400pt}{2.250pt}}
\put(363.67,225){\rule{0.400pt}{3.854pt}}
\multiput(363.17,233.00)(1.000,-8.000){2}{\rule{0.400pt}{1.927pt}}
\put(364.67,225){\rule{0.400pt}{13.731pt}}
\multiput(364.17,225.00)(1.000,28.500){2}{\rule{0.400pt}{6.866pt}}
\put(365.67,175){\rule{0.400pt}{25.776pt}}
\multiput(365.17,228.50)(1.000,-53.500){2}{\rule{0.400pt}{12.888pt}}
\put(366.67,175){\rule{0.400pt}{7.468pt}}
\multiput(366.17,175.00)(1.000,15.500){2}{\rule{0.400pt}{3.734pt}}
\put(367.67,181){\rule{0.400pt}{6.023pt}}
\multiput(367.17,193.50)(1.000,-12.500){2}{\rule{0.400pt}{3.011pt}}
\put(368.67,181){\rule{0.400pt}{3.132pt}}
\multiput(368.17,181.00)(1.000,6.500){2}{\rule{0.400pt}{1.566pt}}
\put(370.17,113){\rule{0.400pt}{16.300pt}}
\multiput(369.17,160.17)(2.000,-47.169){2}{\rule{0.400pt}{8.150pt}}
\put(371.67,113){\rule{0.400pt}{12.768pt}}
\multiput(371.17,113.00)(1.000,26.500){2}{\rule{0.400pt}{6.384pt}}
\put(372.67,166){\rule{0.400pt}{0.723pt}}
\multiput(372.17,166.00)(1.000,1.500){2}{\rule{0.400pt}{0.361pt}}
\put(262.0,323.0){\usebox{\plotpoint}}
\put(374.67,131){\rule{0.400pt}{9.154pt}}
\multiput(374.17,150.00)(1.000,-19.000){2}{\rule{0.400pt}{4.577pt}}
\put(375.67,131){\rule{0.400pt}{3.614pt}}
\multiput(375.17,131.00)(1.000,7.500){2}{\rule{0.400pt}{1.807pt}}
\put(376.67,142){\rule{0.400pt}{0.964pt}}
\multiput(376.17,144.00)(1.000,-2.000){2}{\rule{0.400pt}{0.482pt}}
\put(378.17,124){\rule{0.400pt}{3.700pt}}
\multiput(377.17,134.32)(2.000,-10.320){2}{\rule{0.400pt}{1.850pt}}
\put(379.67,124){\rule{0.400pt}{2.409pt}}
\multiput(379.17,124.00)(1.000,5.000){2}{\rule{0.400pt}{1.204pt}}
\put(380.67,134){\rule{0.400pt}{6.023pt}}
\multiput(380.17,134.00)(1.000,12.500){2}{\rule{0.400pt}{3.011pt}}
\put(381.67,159){\rule{0.400pt}{1.445pt}}
\multiput(381.17,159.00)(1.000,3.000){2}{\rule{0.400pt}{0.723pt}}
\put(382.67,132){\rule{0.400pt}{7.950pt}}
\multiput(382.17,148.50)(1.000,-16.500){2}{\rule{0.400pt}{3.975pt}}
\put(383.67,132){\rule{0.400pt}{9.877pt}}
\multiput(383.17,132.00)(1.000,20.500){2}{\rule{0.400pt}{4.938pt}}
\put(384.67,145){\rule{0.400pt}{6.745pt}}
\multiput(384.17,159.00)(1.000,-14.000){2}{\rule{0.400pt}{3.373pt}}
\put(386.17,145){\rule{0.400pt}{4.100pt}}
\multiput(385.17,145.00)(2.000,11.490){2}{\rule{0.400pt}{2.050pt}}
\put(387.67,161){\rule{0.400pt}{0.964pt}}
\multiput(387.17,163.00)(1.000,-2.000){2}{\rule{0.400pt}{0.482pt}}
\put(388.67,161){\rule{0.400pt}{13.250pt}}
\multiput(388.17,161.00)(1.000,27.500){2}{\rule{0.400pt}{6.625pt}}
\put(389.67,201){\rule{0.400pt}{3.614pt}}
\multiput(389.17,208.50)(1.000,-7.500){2}{\rule{0.400pt}{1.807pt}}
\put(390.67,201){\rule{0.400pt}{0.964pt}}
\multiput(390.17,201.00)(1.000,2.000){2}{\rule{0.400pt}{0.482pt}}
\put(391.67,180){\rule{0.400pt}{6.023pt}}
\multiput(391.17,192.50)(1.000,-12.500){2}{\rule{0.400pt}{3.011pt}}
\put(392.67,180){\rule{0.400pt}{4.818pt}}
\multiput(392.17,180.00)(1.000,10.000){2}{\rule{0.400pt}{2.409pt}}
\put(394,199.67){\rule{0.482pt}{0.400pt}}
\multiput(394.00,199.17)(1.000,1.000){2}{\rule{0.241pt}{0.400pt}}
\put(395.67,189){\rule{0.400pt}{2.891pt}}
\multiput(395.17,195.00)(1.000,-6.000){2}{\rule{0.400pt}{1.445pt}}
\put(396.67,189){\rule{0.400pt}{3.373pt}}
\multiput(396.17,189.00)(1.000,7.000){2}{\rule{0.400pt}{1.686pt}}
\put(397.67,203){\rule{0.400pt}{3.854pt}}
\multiput(397.17,203.00)(1.000,8.000){2}{\rule{0.400pt}{1.927pt}}
\put(398.67,196){\rule{0.400pt}{5.541pt}}
\multiput(398.17,207.50)(1.000,-11.500){2}{\rule{0.400pt}{2.770pt}}
\put(400,194.67){\rule{0.241pt}{0.400pt}}
\multiput(400.00,195.17)(0.500,-1.000){2}{\rule{0.120pt}{0.400pt}}
\put(400.67,181){\rule{0.400pt}{3.373pt}}
\multiput(400.17,188.00)(1.000,-7.000){2}{\rule{0.400pt}{1.686pt}}
\put(402.17,181){\rule{0.400pt}{7.500pt}}
\multiput(401.17,181.00)(2.000,21.433){2}{\rule{0.400pt}{3.750pt}}
\put(403.67,216){\rule{0.400pt}{0.482pt}}
\multiput(403.17,217.00)(1.000,-1.000){2}{\rule{0.400pt}{0.241pt}}
\put(404.67,216){\rule{0.400pt}{5.300pt}}
\multiput(404.17,216.00)(1.000,11.000){2}{\rule{0.400pt}{2.650pt}}
\put(405.67,225){\rule{0.400pt}{3.132pt}}
\multiput(405.17,231.50)(1.000,-6.500){2}{\rule{0.400pt}{1.566pt}}
\put(406.67,225){\rule{0.400pt}{5.300pt}}
\multiput(406.17,225.00)(1.000,11.000){2}{\rule{0.400pt}{2.650pt}}
\put(407.67,237){\rule{0.400pt}{2.409pt}}
\multiput(407.17,242.00)(1.000,-5.000){2}{\rule{0.400pt}{1.204pt}}
\put(408.67,216){\rule{0.400pt}{5.059pt}}
\multiput(408.17,226.50)(1.000,-10.500){2}{\rule{0.400pt}{2.529pt}}
\put(374.0,169.0){\usebox{\plotpoint}}
\put(411.67,216){\rule{0.400pt}{3.132pt}}
\multiput(411.17,216.00)(1.000,6.500){2}{\rule{0.400pt}{1.566pt}}
\put(412.67,224){\rule{0.400pt}{1.204pt}}
\multiput(412.17,226.50)(1.000,-2.500){2}{\rule{0.400pt}{0.602pt}}
\put(413.67,224){\rule{0.400pt}{1.927pt}}
\multiput(413.17,224.00)(1.000,4.000){2}{\rule{0.400pt}{0.964pt}}
\put(415,230.67){\rule{0.241pt}{0.400pt}}
\multiput(415.00,231.17)(0.500,-1.000){2}{\rule{0.120pt}{0.400pt}}
\put(415.67,231){\rule{0.400pt}{5.782pt}}
\multiput(415.17,231.00)(1.000,12.000){2}{\rule{0.400pt}{2.891pt}}
\put(416.67,236){\rule{0.400pt}{4.577pt}}
\multiput(416.17,245.50)(1.000,-9.500){2}{\rule{0.400pt}{2.289pt}}
\put(418.17,236){\rule{0.400pt}{1.900pt}}
\multiput(417.17,236.00)(2.000,5.056){2}{\rule{0.400pt}{0.950pt}}
\put(419.67,239){\rule{0.400pt}{1.445pt}}
\multiput(419.17,242.00)(1.000,-3.000){2}{\rule{0.400pt}{0.723pt}}
\put(420.67,239){\rule{0.400pt}{3.132pt}}
\multiput(420.17,239.00)(1.000,6.500){2}{\rule{0.400pt}{1.566pt}}
\put(421.67,252){\rule{0.400pt}{0.723pt}}
\multiput(421.17,252.00)(1.000,1.500){2}{\rule{0.400pt}{0.361pt}}
\put(422.67,252){\rule{0.400pt}{0.723pt}}
\multiput(422.17,253.50)(1.000,-1.500){2}{\rule{0.400pt}{0.361pt}}
\put(423.67,252){\rule{0.400pt}{0.964pt}}
\multiput(423.17,252.00)(1.000,2.000){2}{\rule{0.400pt}{0.482pt}}
\put(424.67,243){\rule{0.400pt}{3.132pt}}
\multiput(424.17,249.50)(1.000,-6.500){2}{\rule{0.400pt}{1.566pt}}
\put(426,242.67){\rule{0.482pt}{0.400pt}}
\multiput(426.00,242.17)(1.000,1.000){2}{\rule{0.241pt}{0.400pt}}
\put(427.67,228){\rule{0.400pt}{3.854pt}}
\multiput(427.17,236.00)(1.000,-8.000){2}{\rule{0.400pt}{1.927pt}}
\put(428.67,228){\rule{0.400pt}{1.686pt}}
\multiput(428.17,228.00)(1.000,3.500){2}{\rule{0.400pt}{0.843pt}}
\put(430,233.67){\rule{0.241pt}{0.400pt}}
\multiput(430.00,234.17)(0.500,-1.000){2}{\rule{0.120pt}{0.400pt}}
\put(410.0,216.0){\rule[-0.200pt]{0.482pt}{0.400pt}}
\put(431.67,234){\rule{0.400pt}{2.650pt}}
\multiput(431.17,234.00)(1.000,5.500){2}{\rule{0.400pt}{1.325pt}}
\put(432.67,236){\rule{0.400pt}{2.168pt}}
\multiput(432.17,240.50)(1.000,-4.500){2}{\rule{0.400pt}{1.084pt}}
\put(434.17,236){\rule{0.400pt}{3.100pt}}
\multiput(433.17,236.00)(2.000,8.566){2}{\rule{0.400pt}{1.550pt}}
\put(435.67,246){\rule{0.400pt}{1.204pt}}
\multiput(435.17,248.50)(1.000,-2.500){2}{\rule{0.400pt}{0.602pt}}
\put(436.67,246){\rule{0.400pt}{1.686pt}}
\multiput(436.17,246.00)(1.000,3.500){2}{\rule{0.400pt}{0.843pt}}
\put(437.67,249){\rule{0.400pt}{0.964pt}}
\multiput(437.17,251.00)(1.000,-2.000){2}{\rule{0.400pt}{0.482pt}}
\put(438.67,249){\rule{0.400pt}{0.964pt}}
\multiput(438.17,249.00)(1.000,2.000){2}{\rule{0.400pt}{0.482pt}}
\put(431.0,234.0){\usebox{\plotpoint}}
\put(440.67,247){\rule{0.400pt}{1.445pt}}
\multiput(440.17,250.00)(1.000,-3.000){2}{\rule{0.400pt}{0.723pt}}
\put(442.17,247){\rule{0.400pt}{1.300pt}}
\multiput(441.17,247.00)(2.000,3.302){2}{\rule{0.400pt}{0.650pt}}
\put(443.67,245){\rule{0.400pt}{1.927pt}}
\multiput(443.17,249.00)(1.000,-4.000){2}{\rule{0.400pt}{0.964pt}}
\put(445,244.67){\rule{0.241pt}{0.400pt}}
\multiput(445.00,244.17)(0.500,1.000){2}{\rule{0.120pt}{0.400pt}}
\put(445.67,240){\rule{0.400pt}{1.445pt}}
\multiput(445.17,243.00)(1.000,-3.000){2}{\rule{0.400pt}{0.723pt}}
\put(446.67,240){\rule{0.400pt}{0.723pt}}
\multiput(446.17,240.00)(1.000,1.500){2}{\rule{0.400pt}{0.361pt}}
\put(447.67,241){\rule{0.400pt}{0.482pt}}
\multiput(447.17,242.00)(1.000,-1.000){2}{\rule{0.400pt}{0.241pt}}
\put(449,240.67){\rule{0.241pt}{0.400pt}}
\multiput(449.00,240.17)(0.500,1.000){2}{\rule{0.120pt}{0.400pt}}
\put(450.17,242){\rule{0.400pt}{1.100pt}}
\multiput(449.17,242.00)(2.000,2.717){2}{\rule{0.400pt}{0.550pt}}
\put(451.67,243){\rule{0.400pt}{0.964pt}}
\multiput(451.17,245.00)(1.000,-2.000){2}{\rule{0.400pt}{0.482pt}}
\put(452.67,243){\rule{0.400pt}{0.964pt}}
\multiput(452.17,243.00)(1.000,2.000){2}{\rule{0.400pt}{0.482pt}}
\put(454,245.67){\rule{0.241pt}{0.400pt}}
\multiput(454.00,246.17)(0.500,-1.000){2}{\rule{0.120pt}{0.400pt}}
\put(454.67,246){\rule{0.400pt}{0.482pt}}
\multiput(454.17,246.00)(1.000,1.000){2}{\rule{0.400pt}{0.241pt}}
\put(456,246.67){\rule{0.241pt}{0.400pt}}
\multiput(456.00,247.17)(0.500,-1.000){2}{\rule{0.120pt}{0.400pt}}
\put(457,246.67){\rule{0.241pt}{0.400pt}}
\multiput(457.00,246.17)(0.500,1.000){2}{\rule{0.120pt}{0.400pt}}
\put(440.0,253.0){\usebox{\plotpoint}}
\put(459.67,246){\rule{0.400pt}{0.482pt}}
\multiput(459.17,247.00)(1.000,-1.000){2}{\rule{0.400pt}{0.241pt}}
\put(460.67,246){\rule{0.400pt}{0.482pt}}
\multiput(460.17,246.00)(1.000,1.000){2}{\rule{0.400pt}{0.241pt}}
\put(461.67,246){\rule{0.400pt}{0.482pt}}
\multiput(461.17,247.00)(1.000,-1.000){2}{\rule{0.400pt}{0.241pt}}
\put(458.0,248.0){\rule[-0.200pt]{0.482pt}{0.400pt}}
\put(464,244.67){\rule{0.241pt}{0.400pt}}
\multiput(464.00,245.17)(0.500,-1.000){2}{\rule{0.120pt}{0.400pt}}
\put(463.0,246.0){\usebox{\plotpoint}}
\put(469,244.67){\rule{0.241pt}{0.400pt}}
\multiput(469.00,244.17)(0.500,1.000){2}{\rule{0.120pt}{0.400pt}}
\put(470,244.67){\rule{0.241pt}{0.400pt}}
\multiput(470.00,245.17)(0.500,-1.000){2}{\rule{0.120pt}{0.400pt}}
\put(471,244.67){\rule{0.241pt}{0.400pt}}
\multiput(471.00,244.17)(0.500,1.000){2}{\rule{0.120pt}{0.400pt}}
\put(465.0,245.0){\rule[-0.200pt]{0.964pt}{0.400pt}}
\put(472.0,246.0){\rule[-0.200pt]{7.468pt}{0.400pt}}
\end{picture}